%% file: main.tex
\documentclass[letterpaper,12pt]{article}

\input{sections/preamble} 

\begin{document}


\input{sections/coverpage} 

\thispagestyle{empty} 







\input{sections/content}

\newpage
\singlespacing


    \bibliographystyle{unsrt} 
	\bibliography{main} 

















\newpage




\end{document}

%% file: sections/preamble.tex
    \usepackage{abstract} 
    
    \usepackage{titlesec} 

    \usepackage{authblk} 

	\usepackage{datetime} 
    \newdateformat{usvardate}{
  	\monthname[\THEMONTH] \ordinal{DAY}, \THEYEAR}

  	\usepackage[bottom]{footmisc} 
    
    \usepackage{fancyhdr}
    \fancyheadoffset{0cm}
    \providecommand{\keywords}[1]{\textbf{\textit{Keywords---}} #1}

\newcommand\wordcount{%
  \immediate\write18{texcount -utf8 -merge -sum -incbib -dir -sub=none -brief \jobname.tex | cut -d : -f 1 > 'count.txt'}%
  \input{count.txt}\ignorespaces words%
}

    \usepackage{amssymb}
    \usepackage{amsthm}
    \usepackage{newtxmath}

    \usepackage{microtype} 
    \usepackage[utf8]{inputenc}
    \usepackage{newtxtext} 
    \usepackage[misc]{ifsym}
  
	\usepackage{setspace} 
    \usepackage{lineno,xcolor} 	

	\usepackage[top=1.5in, bottom=1.5in, left=1in, right=1in]{geometry}

	\usepackage[colorinlistoftodos]{todonotes} 
	\usepackage{soul} 
    
    \usepackage{float} 
    
    \usepackage{printlen}
    \usepackage{subcaption}
    \usepackage{tcolorbox}

    \usepackage{longtable} 
    \usepackage{tabularx} 
    \newcolumntype{L}[1]{>{\raggedright\arraybackslash}p{#1}}
    \newcolumntype{C}[1]{>{\centering\arraybackslash}p{#1}}
    \newcolumntype{R}[1]{>{\raggedleft\arraybackslash}p{#1}}
    \usepackage{relsize} 

    \usepackage{cite}
	\usepackage{hyperref} 

%% file: sections/coverpage.tex
\newgeometry{top=1.7cm, bottom=2cm, left=2.5cm, right=2.5cm} 

\begin{titlepage}

  \newcommand{\HRule}{\rule{\linewidth}{0.5mm}} 

  \center 


  \begin{minipage}{\textwidth}
    \centering
    \includegraphics[width=0.12\textwidth]{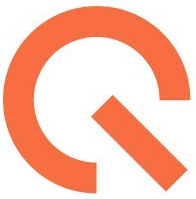}
    \hspace{10pt}
    \includegraphics[width=0.12\textwidth]{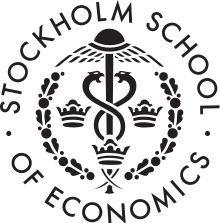}
    \hspace{10pt}
    \includegraphics[width=0.12\textwidth]{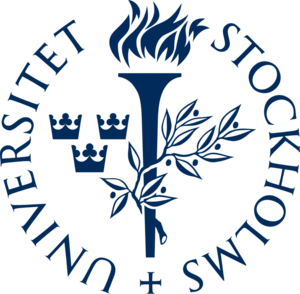}
  \end{minipage}
  \\[1cm]




  \HRule \\[0.5cm]
  { \huge \bfseries Using Deep Learning to Find the Next Unicorn: A Practical Synthesis} 
  \\[0.1cm]
  \HRule \\[1.0cm]
 

  \vspace{.2 cm}
  Lele \textsc{Cao}\, \textsuperscript{\Letter}\\
  Motherbrain, EQT\\
  \texttt{lele.cao@eqtpartners.com\;\;\;\;\;\;\;\;caolele@gmail.com}\\
  \vspace{.5 cm}
  Vilhelm \textsc{von Ehrenheim}\\
  Motherbrain, EQT\\
  \texttt{vilhelm.vonehrenheim@eqtpartners.com}\\
  \vspace{.5 cm}
  Sebastian \textsc{Krakowski}\\
  House of Innovation, Stockholm School of Economics\\
  \texttt{sebastian.krakowski@hhs.se}\\
  \vspace{.5 cm}
  Xiaoxue \textsc{Li}\\
  Department of Political Science, Stockholm University \\
  \texttt{xiaoxue.li@statsvet.su.se} \\
  \vspace{.5 cm}
  Alexandra \textsc{Lutz}\\
  Motherbrain, EQT\\
  \texttt{alexandra.lutz@eqtpartners.com}


\vspace{1 cm}

  \begin{flushleft}
\begin{tcolorbox}[colframe=blue]
\footnotesize
\raisebox{-0.1\height}{\includegraphics[height=1.2em]{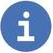}}
A condensed version \cite{cao-etal-2023-using} is peer reviewed and published by IJCAI 2023 (The 32nd International Joint Conference on Artificial Intelligence) Workshop: \url{https://aclanthology.org/2023.finnlp-1.6}.

\vspace{1em}
Chicago Citation Format:

\texttt{\scriptsize Cao, Lele, Vilhelm von Ehrenheim, Sebastian Stan, Xiaoxue Li, and Alexandra Lutz.~"\textbf{Using Deep Learning to Find the Next Unicorn:~A Practical Synthesis on Optimization Target, Feature Selection, Data Split and Evaluation Strategy.}" \textit{Proceedings of the IJCAI Joint Workshop on the 5th Financial Technology and Natural Language Processing (FinNLP) and the 2nd Multimodal AI for Financial Forecasting (Muffin)}, pp.~63-73, 2023.}
\end{tcolorbox}
\end{flushleft}

  \vspace{0.8 cm}
  \begin{flushleft}
  \footnotesize
\text{\Letter}\, Please send correspondence to the first author -- Lele Cao, Motherbrain AI Research, EQT Group, Regeringsgatan 25, 11153 Stockholm, Sweden; e-mail: caolele@gmail.com. 
\end{flushleft}

\vfill 

\end{titlepage}

\restoregeometry 


\newpage

%% file: sections/content.tex

\vspace{1.5 cm}
\doublespacing
\begin{abstract}
\noindent Startups often represent newly established business models associated with disruptive innovation and high scalability. They are commonly regarded as powerful engines for economic and social development. Meanwhile, startups are heavily constrained by many factors such as limited financial funding and human resources. Therefore, the chance for a startup to eventually succeed is as rare as ``spotting a unicorn in the wild''. Venture Capital (VC) strives to identify and invest in unicorn startups during their early stages, hoping to gain a high return. To avoid entirely relying on human domain expertise and intuition, investors usually employ data-driven approaches to forecast the success probability of startups. Over the past two decades, the industry has gone through a paradigm shift moving from conventional statistical approaches towards becoming machine-learning (ML) based. Notably, the rapid growth of data volume and variety is quickly ushering in deep learning (DL), a subset of ML, as a potentially superior approach in terms of capacity and expressivity. In this work, we carry out a literature review and synthesis on DL-based approaches, covering the entire DL life cycle. The objective is a) to obtain a thorough and in-depth understanding of the methodologies for startup evaluation using DL, and b) to distil valuable and actionable learning for practitioners. To the best of our knowledge, our work is the first of this kind. 
\end{abstract}

\vspace{2 cm}
\begin{flushleft}
\keywords{Startup, Success Prediction, Unicorn, Deep Learning, Machine Learning, Venture Capital, Investment, Big Data, Practical Synthesis}
\end{flushleft}

\newpage


\doublespacing

\section{Introduction}

A ``startup'' has many variants of definitions; up until this date there is no consensus on the standard definition.
Santisteban and Mauricio \cite{j8_santisteban2021critical} synthesized many popular definitions, and discovered some common labels such as ``new'', ``small'',  ``rapid growth'', ``high risk'', 
where ``small'' is often approximated by limited financial funds and human resources \cite{j22_skawinska2020success}. 
Much of the literature, e.g., \cite{blank2013lean}, associates startups with disruptive innovation and high scalability.
As a result, 
\begin{quote}
``{\it A startup is a dynamic, flexible, high risk, and recently established company that typically represents a reproducible and scalable business model. It provides innovative products and/or services, and has limited financial funds and human resources.}'' \cite{j8_santisteban2021critical,j22_skawinska2020success,blank2013lean}
\end{quote}

Since startups stimulate growth, generate jobs and tax revenues, and promote many other socioeconomically beneficial factors \cite{acs2007entrepreneurship},
they are commonly regarded as powerful engines for economic and social development, especially after economic, environment, and epidemic crisis such as COVID-19\footnote{Coronavirus disease 2019 (COVID-19) is a contagious disease caused by the severe acute respiratory syndrome coronavirus 2 (SARS-CoV-2). The first case was identified in 2019.} \cite{c1_zhang2021scalable}.
As the startups continue to develop, they often increasingly rely on external funds (as opposed to internal funds from founders and co-founders), from either domestic or foreign capital markets, to unlock a high rate of growth that usually corresponds to a ``hockey stick'' growth curve (i.e. a linear line on a log scale) \cite{marmer2011startup}.

Startups may receive funds from multiple sources like Venture Capital (VC) and debt financing; up till this date, the dominating source has been VC.
As an industry, VC seeks opportunities to invest in startups with great potential (in the sense of financial returns) to grow and successfully exit.
The risk-return trade-off tells us that the potential return rises with a corresponding increase in risk\footnote{Statistics revealing the high risk of funding startups: on average, only around 60\% of the startups survive for more than 3 years since founded \cite{hyytinen2015does}; top 2\% of VC funds receive 95\% of the returns in the entire industry \cite{j26_bai2021startup}; VC typically has only 10\% rate of achieving an ROI (return on investment) of 100\% or more \cite{shane2012importance,t3_Unal2019Machine}.}.
As a consequence, VC firms usually strive to mitigate this risk by improving their 1) {\it deal sourcing}\footnote{Deal sourcing is the process by which investors identify investment opportunities.} and screening and 2) {\it value-add} process \cite{teten2013lower}.
In this survey, we will focus on the published work around the former approach, i.e. finding the {\it startup unicorn}\footnote{
Unicorn and near-unicorn startups are private, venture-backed firms with a valuation of at least \$500 million at some point \cite{chernenko2021mutual}.
} as accurately as possible during the deal sourcing phase.

Finding the unicorn from candidate startups is a complex task with great uncertainty because of many factors such as vague and prone-to-change business ideas, no proof-of-concept prototype when applicable, no organic revenue.
This creates a low information situation, where VC firms often have to make investment decisions based on insufficient information (e.g. lack of financial data) \cite{c9_dellermann2021finding}.
Therefore a VC's deal sourcing process traditionally turns out to be manual and empirical,
leaving estimations of the ROI (return on investment) heavily dependent on the human investors' decisions.
As pointed out in \cite{cumming2010local}, human investors are inherently biased and intuition alone cannot consistently drive good decisions.
A better approach should leverage big data to 
\begin{itemize}
\item debias the decisions, so that the individual investment decision made for a particular startup is expected to drive lower risk and higher ROI;
\item enable automation, so that more startups can be evaluated without requesting extra amount of time.
\end{itemize}
To that end, over the past two decades, data driven approaches have been dominating the research around {\it startup success prediction} (i.e. identifying startups that eventually turn into unicorns).
However, the majority is analytical and statistical as opposed to ML (machine learning) approaches.
Conventional statistical work (e.g. \cite{lussier2001crossnational,davila2003venture,j19_hochberg2007whom,j34_nahata2008venture,lussier2010three,samila2011venture,puri2012life,nanda2013investment,okrah2018exploring,islam2018signaling,t7_saini2018picking,prohorovs2019startup,j33_malmstrom2020they,gompers2020venture,j30_kaiser2020value,rj1_pasayat2020factors,diaz2021econometric,j8_santisteban2021critical,t10_melnychuk2021approved}) mostly starts with defining some hypotheses\footnote{Hypothesis often assumes certain impact of some factors to startup success. 
For example ``the founder's past entrepreneurial experience influences the likelihood of success \cite{diaz2021econometric}''.}, followed by testing them using statistical tools; 
the outcome of these work is often conclusions around correlation and/or causality between some factors and the success likelihood of startups.

\begin{figure}[!t]
\centerline{\includegraphics[width=.7\linewidth]{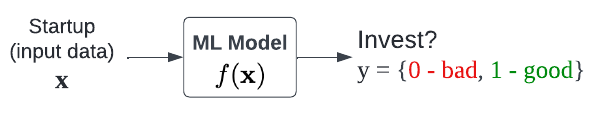}}
\vspace{-8pt}
\caption{\label{fig:ml-approach}\textbf{High-level overview of ML (machine learning) based startup sourcing}\newline The ML model is trained to approximate a function $f(\cdot)$ so that the input data $\mathbf{x}$ describing a startup can be mapped to an output variable $y$ indicating the recommended investment propensity that can be either discrete (good vs. bad) or continuous (success probability).}
\end{figure}

In conventional statistical research, good research hypotheses need to be simple, concise, precise, testable;
and most importantly, they should be grounded in past knowledge, gained from the literature review or from theory \cite{williamson2002research}.
Therefore, it is not a easy task to come up with good hypotheses.
Over the last few years, researchers have started investigating the possibility to perform {\it hypothesis mining} from data using ML algorithms to avoid manually defining hypotheses upfront.
Hypothesis mining aims to summarize (instead of manually define) hypotheses by carrying out explainability analysis (cf. Section~\ref{sec:explainability}) on the trained ML models \cite{w5_guerzoni2019survival}.
For example, with a labeled (i.e. knowing which startups eventually become unicorns) dataset containing many attributes for many companies;
one can directly start off with training an ML model to predict unicorns (i.e. prediction target) using the entire dataset (all companies and attributes).
By explaining and quantifying how the change of certain attributes would change the prediction target, one may distil hypothesis that describes the relation between the attributes in scope and the prediction target.
In comparison to exploratory data analysis, hypothesis mining is a much more structured procedure that trains an ML model using the entire dataset at hand.
As illustrated in Figure~\ref{fig:ml-approach}, the ML-based approaches \cite{c5_xiang2012supervised,j36_rouhani2013erp,j27_liang2016predicting,zhong2016or,krishna2016predicting,bohm2017business,zhong2018startup,t5_bento2018predicting,arroyo2019assessment,j15_shin2019network,t6_unal2019searching,w5_guerzoni2019survival,li2020prediction,sadatrasoul2020hybrid,j17_bonaventura2020predicting,kipkogei2021tree,veloso2020predicting,cavicchioli2021learning,zbikowski2021machine,t8_kamal2021modeling,singhal2022data} require practitioners to define the input data $\mathbf{x}$ and annotation $y$ (labeling good or bad investment according to some criteria) before training a model $f(\cdot)$ that maps $\mathbf{x}$ to $y$, i.e. $y=f(\mathbf{x})$.
There are already a few survey papers \cite{rj1_pasayat2020factors,bargagli2021supervised} about ML-based work.

\begin{figure}[t!]
\centering
\subcaptionbox{\footnotesize ANN with one hidden layer.\label{fig:non-dl-ann}}
{\includegraphics[height=4cm]{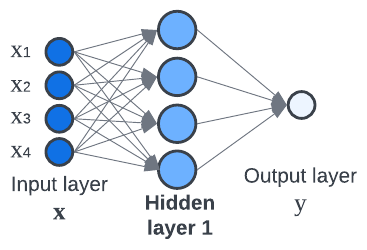}}
\hspace{1.0cm}
\subcaptionbox{\footnotesize ANN with more than one hidden layer.\label{fig:dl-ann}}
{\includegraphics[height=4cm]{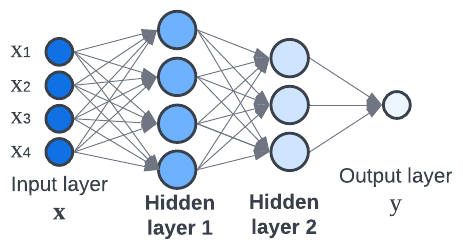}}
\caption{\label{fig:anns}\textbf{DL (deep learning) utilizes ANNs (artificial neural networks) with at least two hidden layers; thus (a) is not considered as a DL model in this work.}\newline The input data $\mathbf{x}$ is fed into the input layer before flowing through the hidden layers. The output layer generates the final prediction $y$. The connections (fully or partly connected) between the adjacent layers carry trainable weights.}
\end{figure}

With the rapid growth of dataset size and diversity (origin and modality), traditional ML models\footnote{Traditional ML models that are frequently applied: decision tree (e.g. \cite{arroyo2019assessment}), random forest (e.g. \cite{krishna2016predicting}), logistic regression (e.g. \cite{kipkogei2021tree}), gradient boosting (e.g. \cite{zbikowski2021machine}), SVM (support vector machine, e.g. \cite{j27_liang2016predicting}), $k$-means clustering (e.g. \cite{cavicchioli2021learning}), Bayesian network (e.g. \cite{c5_xiang2012supervised}).} sometimes struggle to directly and fully fit the big-and-raw dataset due to lack of model {\it capacity} and {\it expressivity}\footnote{\label{footnote:capacity-expressivity}{\it Expressivity} describes the classes of functions a model can approximate, while {\it capacity} measures how much ``brute force'' ability the model has to fit the data.}.
Most recently, DL (deep learning) algorithms caught the eyes of increasing number of researchers hunting for unicorns. 
DL, by definition, represents a subset of ML methods \cite{lecun2015deep}, and is implemented (entirely or partly) with ANNs (artificial neural networks) that utilize at least two hidden layers of neurons as shown in Figure~\ref{fig:dl-ann}.
The {\it capacity} of DL can be controlled by the number of neurons (width) and layers (depth) \cite{goodfellow2016deep}.
Deep ANNs are exponentially {\it expressive} with respect to their depth \cite{raghu2017expressive}, making ANNs universal function approximators\footnote{Simply speaking, ANN containing a specific number of neurons in the hidden layer(s) can approximate almost any known function.} \cite{hornik1989multilayer}.
As a well-known international investment firm practicing data-driven approaches to find startup unicorns,
we strive to 
\begin{itemize}
\item obtain a thorough and in-depth understanding of the methodologies for startup evaluation using DL, and 
\item distil important and actionable learning for practitioners in this domain.
\end{itemize}
To achieve these goals, we carry out a comprehensive literature survey on using DL to evaluate startups (mostly for investment deal sourcing) \cite{c23_kim2017does,c20_lee2018content,c22_yu2018prediction,c3_sharchilev2018web,w3_gastaud2019varying,c21_cheng2019success,j46_kim2020recommendation,c6_ghassemi2020automated,j41_yeh2020machine,j42_srinivasan2020ensemble,j44_kaminski2020predicting,c9_dellermann2021finding,c15_ferrati2021deep,c19_garkavenko2021valuation,c16_chen2021trend,j2_ross2021capitalvx,c1_zhang2021scalable,j26_bai2021startup,j37_kinne2021predicting,j43_shi2021leveraging,t1_stahl2021leveraging,t2_horn2021deep,w2_yin2021solving,w1_lyu2021graph,j16_allu2022predicting,j9_tang2022deep,b1_ang2022using,j47_wu2022estimating,w6_garkavenko2022you}\footnote{The literature include 29 peer-reviewed English papers/theses addressing startup success prediction using DL methods. We do not apply any restrictions on publication years, geo-location of study, or publication type. The papers/theses are sourced using a combination of three approaches: (1) recommendation by investment professionals and researchers, 
(2) keywords searching in Google Search (\protect\url{google.com}), Google Scholar (\protect\url{scholar.google.com}), IEEE (\protect\url{ieee.org}), ACM (\protect\url{acm.org}), Scopus (\protect\url{scopus.com}), Wiley (\protect\url{wiley.com}), Springer (\protect\url{springer.com}) and Web of Science (\protect\url{clarivate.com}); 
(3) cross reference among papers/theses.}.
To the best of our knowledge till this date, our work is the first of this kind.
According to our high-level synthesis, most DL based approaches comprise nine consecutive key tasks listed below.
\begin{itemize}
\item {\it Define the prediction problem}, i.e. what specific question do we expect the model to answer?
\item {\it Define the startup success criteria}, so that the data can be annotated accordingly to train the model.
\item {\it Gather the data} to be used as model input; it is inevitable to make decisions around the source, category, modality and size of the data.
\item {\it Process the data}, e.g. normalize, augment, debias, balance and densify the data to drive better performance of DL models.
\item {\it Split the data}, i.e. divide the dataset into on-overlapping training, evaluation and testing sets. 
\item {\it Select the DL model variants}: there are many DL model variants, thus one can only pick a limited number of variants for experimentation.
\item {\it Evaluate the trained DL model}: the performance of each trained (using the training set) DL model is evaluated over the evaluation set to find the best hyper-parameters.
\item {\it Explain the predictions from the trained DL model}: DL models are commonly regarded as ``black-box'' models, hence explainability is required to promoting transparency, build trust, and capture feedback.
\item {\it Deploy the trained DL model}: in applied scenarios, the trained DL model with satisfying test results will be productized\footnote{Model productization, a.k.a. model deployment, is the procedure by which practitioners integrate a machine learning model into an existing production environment to make practical business decisions based on input data. It is one of the last stages in the DL engineering life cycle and can be one of the most cumbersome.} to answer investors' questions on demand.
\end{itemize}
In the rest of this paper, we will consecutively discuss the key aspects to accomplish these nine tasks (corresponds to nine sections that follows). 
In each section, we present a literature synthesis and practical learnings to facilitate a successful application of DL to ``unicorn hunting''.
The practical learning is synthesized from both the literature and our industrial experiences.
For the sake of clarity, the headline of each section is made concrete to reflect the key learning.

\section{Avoid Predicting Success and Compatibility Simultaneously}
\label{sec:success-compatibility}
\begin{figure}[t!]
\centering
\subcaptionbox{\footnotesize Ideal DL model outputs.\label{fig:dl-approach-split-output}}
{\includegraphics[height=2.07cm]{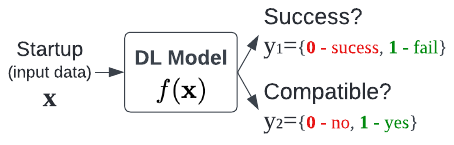}}
\subcaptionbox{\footnotesize Apply compatibility filtering before prediction.\label{fig:dl-approach-filter}}
{\includegraphics[height=2.07cm]{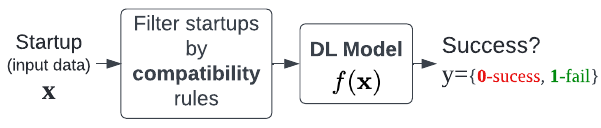}}
\caption{\label{fig:dl-approach-success-compatibility}\textbf{Two ways of addressing the probability of startup success and compatibility.}\newline Most DL-based work do not explicitly consider startup success and compatibility at the same time. Two feasible solutions are presented here. We recommend solution (b) over (a) due to its simplicity, flexibility, and closer approximation to real use cases.}
\end{figure}

The DL model is generally expected to suggest whether funding a certain startup is likely to fulfill the investment goal (cf. Figure~\ref{fig:ml-approach}).
As a matter of fact, the common investment goal usually embodies two components:
\begin{itemize}
\item Success: the startup-in-scope eventually achieves a {\it successful outcome} after the decision of investing;
and the different ways of defining that {\it successful outcome} will be introduced in Section~\ref{sec:success_criteria}. 
\item Compatibility: the startup matches the preference of the investment intention, which originates from many practical requirements such as geographical emphasis, sector focus, portfolio conflict, investment mandate, and exit opportunities \cite{j46_kim2020recommendation}.
\end{itemize}
The vast majority of the surveyed research does not explicitly distinguish predicting the success and compatibility; 
they either implicitly address both simultaneously (e.g.~\cite{c3_sharchilev2018web}), or simply ignore the compatibility part (e.g.~\cite{c15_ferrati2021deep}).
In practice, VC investors are restricted to funding startups that meet the compatibility requirement. Ideally, we wish to split the output in Figure~\ref{fig:ml-approach} into two outputs as shown in Figure~\ref{fig:dl-approach-split-output}.

However, training one single DL model to predict both success and compatibility is more challenging than predicting merely one of them.
Moreover, when engineering an applied solution\footnote{There are a few VC funds who try to develop and commercialize DL-based startup evaluation models, such as the Motherbrain platform from EQT Ventures (\url{eqtventures.com}) \cite{b2_corea2019ai}.} to facilitate VC operations, we observe that the compatibility definition is prone to change when time, context, or the actual user changes.
Inspired by \cite{j46_kim2020recommendation,c16_chen2021trend,j47_wu2022estimating}, we propose to perform {\it compatibility filtering} before the DL model prediction, as demonstrated in Figure~\ref{fig:dl-approach-filter}.
The compatibility filtering essentially removes the startups that are regarded as incompatible with the preference of investment professionals or the fund specifications, resulting in a set of ``feasible'' startups for consideration.
This is mostly achieved by defining some filters using factors like geo-location (e.g. EU and US), business sector (e.g. Fintech and Biotech), customer focus (e.g. B2B/B2C/B2B2C)\footnote{B2B: business-to-business. B2C: business-to-consumer. B2B2C: business-to-business-to-consumer, where businesses access customers through a third party.}, and development stage (e.g. approximated by the total funding received by the startup in scope).

Applying compatibility filtering before DL model (cf.~Figure~\ref{fig:dl-approach-filter}) simplifies the DL model, makes it flexible to use, and adapts the model better to investment preferences. 
An additional benefit of doing so is that a model trained on a subset of successful startups relevant to a given VC fund will, by nature, produce more relevant predictions for that VC fund.

\section{Clearly Define the Success Criteria of Startups}
\label{sec:success_criteria}
\begin{figure}[t!]
\centering
\subcaptionbox{\footnotesize Distribution of the adopted startup success criteria.\label{fig:success-criteria-dist}}
{\includegraphics[height=6cm]{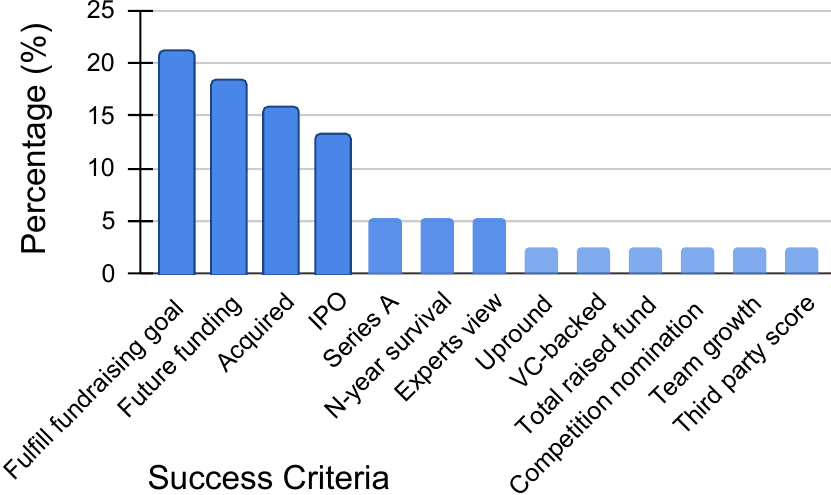}}
\hspace{2cm}
\subcaptionbox{\footnotesize Criteria mixture.\label{fig:success-criteria-mixture}}
{\includegraphics[height=6cm]{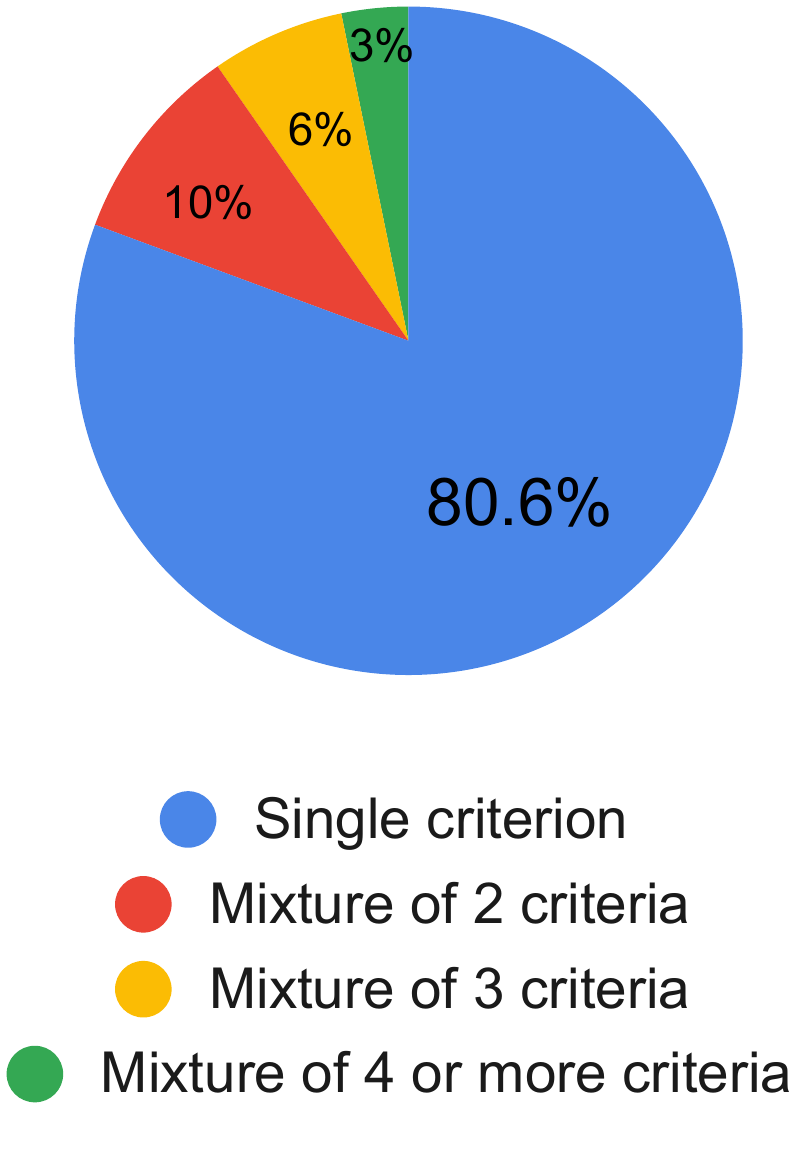}}
\caption{\label{fig:success-criteria-dist-mixture}\textbf{Summary of the adopted criteria to evaluate startup success.}\newline  (a) shows the percentage of each success criterion sorted by the their occurrences. (b) shows the percentage of combining different number of criteria together.}
\end{figure}

Identifying potential unicorns relies on accurate prediction of startup success. 
So far there is no universally agreed definition of "true success"; most of the existing definitions commonly focus on ``growth'' which can be measured from different perspectives like revenue, employees, and valuation, to name a few.
We summarize the adopted definitions from the reviewed literature \cite{c23_kim2017does,c20_lee2018content,c22_yu2018prediction,c3_sharchilev2018web,w3_gastaud2019varying,c21_cheng2019success,j46_kim2020recommendation,c6_ghassemi2020automated,j41_yeh2020machine,j42_srinivasan2020ensemble,j44_kaminski2020predicting,c9_dellermann2021finding,c15_ferrati2021deep,c19_garkavenko2021valuation,c16_chen2021trend,j2_ross2021capitalvx,c1_zhang2021scalable,j26_bai2021startup,j37_kinne2021predicting,j43_shi2021leveraging,t1_stahl2021leveraging,t2_horn2021deep,w2_yin2021solving,w1_lyu2021graph,j16_allu2022predicting,j9_tang2022deep,b1_ang2022using,j47_wu2022estimating,w6_garkavenko2022you} in Figure~\ref{fig:success-criteria-dist-mixture}, showing each criterion's popularity among researchers. 
All {\it success criteria} are quantities in relation to a predefined duration since the time point of evaluation.

\begin{enumerate}
\item {\bf Fulfill the preset fundraising goal} \cite{c20_lee2018content,c22_yu2018prediction,c21_cheng2019success,j41_yeh2020machine,j42_srinivasan2020ensemble,j43_shi2021leveraging,j44_kaminski2020predicting,j47_wu2022estimating,j9_tang2022deep}: the goal (the expected amount of money) of the fund-raise campaign or plan is reached or surpassed, which is common among crowdfunding projects. The readers should be cautious not to confuse with the fund-raise goal of investors.
\item {\bf Future funding} \cite{c3_sharchilev2018web,w3_gastaud2019varying,c16_chen2021trend,j2_ross2021capitalvx,t1_stahl2021leveraging,w2_yin2021solving,w6_garkavenko2022you}: any future funding raised above a low-bar amount. 
\item {\bf Acquired} \cite{b1_ang2022using,c15_ferrati2021deep,j2_ross2021capitalvx,j46_kim2020recommendation,w1_lyu2021graph,w2_yin2021solving}: one company purchases and takes over the operations and assets of the startup.
\item {\bf IPO} (initial public offering) \cite{b1_ang2022using,c15_ferrati2021deep,j2_ross2021capitalvx,w1_lyu2021graph,w2_yin2021solving}: it offers shares to the public in a new stock issuance for the first time; IPO allows the company to raise equity capital from public investors.
\item {\bf Series A} \cite{c1_zhang2021scalable,c9_dellermann2021finding}: the startup receives the first VC funding round after the seed and angel rounds.
\item {\bf $\boldmath{N}$-year survival} \cite{c6_ghassemi2020automated,j2_ross2021capitalvx}: the firm is still operating after $N$ years.
\item {\bf Experts view} \cite{j26_bai2021startup,j37_kinne2021predicting}: the quantified review from human experts.
\item {\bf Upround} \cite{b1_ang2022using}: the (post-money) valuation after a future funding round is higher than the current valuation.
\item {\bf VC-backed} \cite{c19_garkavenko2021valuation}: the startup is funded by one or more VC firms.
\item {\bf Total raised funding} \cite{c23_kim2017does}: the accumulated amount of funding received (the higher the better), which is often used as a regression target.
\item {\bf Competition nomination} \cite{c6_ghassemi2020automated}: the idea of the startup wins (or nominated by the committee) a entrepreneurial competition.
\item {\bf Team growth} \cite{t2_horn2021deep}: whether the team size has experienced a fast growth or not, such as ``{\small \it a minimum of $x$\% increase from at least 10 initial employees}''.
\item {\bf Third party score} \cite{j16_allu2022predicting}: some data sources provide certain firm evaluation scores, such as the ``trend score'' from Crunchbase\footnote{\url{www.crunchbase.com}}. 
\end{enumerate}

While the first twelve criteria are intuitively sound, we question the effectiveness of the last criterion of taking the 3rd-party (algorithmic) scores as ground truth to train the DL model, because it is guaranteed to obtain a model inferior to the 3rd-party method (often unknown).
Additionally, there is no financial based success criteria\footnote{A few ML-based (instead of DL-based) work \cite{lussier2001crossnational,lussier2010three} have investigated using financial based success criteria.} adopted in the DL-based work, which is a consequence of missing rich operating data \cite{gompers2020venture} before exiting the startup phase and entering the {\it growth phase} \cite{j22_skawinska2020success}.
Although the definition of a successful startup has many versions, for investors, it is relatively straightforward: a profitable exit, often in the form of acquisition or IPO, which incur high ROI \cite{b1_ang2022using}.
However, if you ask an investment professional, short-term events like funding rounds have a higher adoption rate than longer-term acquisition/IPO; the reason is twofold: 1) acquisition/IPO is extremely scarce as very few startups achieve these milestones; and 2) it occurs very late in startup's trajectory, hence potentially weakening the correlation between early data and late success \cite{t1_stahl2021leveraging}. 
Finally, the choice of success criteria also depends on whose perspectives we take.

\begin{itemize}
\item {\bf Investor}'s perspective: the investors' view of success, which typically include high ROI exit events such as acquisition and IPO \cite{w3_gastaud2019varying,b1_ang2022using}.
\item {\bf Founder}'s perspective: what results do founders regard as a successful entrepreneurial outcome? Examples of this kind are competition nomination, series A, and profitable operation \cite{prohorovs2019startup,b1_ang2022using}.
\item {\bf Policy maker}'s perspective: the view of the entities (e.g.,~governments or authorities) who set the plan pursued by a government or business; the policy maker usually considers a broad socio-economic impact, such as job creation and patentability\cite{j30_kaiser2020value,t2_horn2021deep}.
\end{itemize}

In most cases, different success criteria do not conflict with each other, implying the possibility to combine multiple criteria; but this kind of {\it criteria mixture} is still under-investigated as illustrated in Figure~\ref{fig:success-criteria-mixture}.
Generally speaking, one can combine multiple criteria with logical \verb|AND| operators (e.g. \cite{w2_yin2021solving,b1_ang2022using}), or use each criterion separately in a {\it multi-task training} (cf. Section~\ref{sec:model-selection}) setup \cite{j43_shi2021leveraging}. 
It is theoretically proved that multi-task training can reduce the risk of overfitting\footnote{Overfitting happens when a DL model has an overly high capacity so that it is able to fit the entire dataset (including the noise), hence the model performs poorly on unseen data.} \cite{baxter1997bayesian}. 
Based on the discussions above, we highlight some key recommendations from an investor's perspective, when defining the success criteria of startups.

\begin{itemize}
\item Start with interviewing the users and stakeholders of your model to find out their definition of startup success.
\item Avoid adopting success criteria that are mostly valued by non-investors like founders and policy makers.
\item Whenever possible, assemble the success criteria that lead to more annotated (labeled) data, allowing bigger model capacity.
\item Prioritize the short-term event over longer-term ones for the sake of stronger data-label correlation.
\item Consider experimenting a mixture of criteria to incorporate more perspectives and possibly preventing overfitting the DL model.
\end{itemize}

\section{Use Multi-modal, Unstructured, Free and Extrinsic Data}
DL models need data input to make predictions. 
Before we start gathering input data for model, we might be able to benefit from understanding what input(s) humans use to make decisions.
When investment professionals (i.e.~humans) try to forecast the success of early stage startups, they make use of two cognitive modes: {\it intuitive} and {\it analytical}.
The {\it intuitive mode} is characterized by processing ``soft'' signals (e.g. innovativeness and personality of entrepreneur) that are mostly {\it qualitative}; 
and humans are still the ``golden standard'' for this mode \cite{baer2014gold}.
The {\it analytical mode}, on the other hand, deals with ``hard'' facts (e.g. industry and team size) that are often {\it quantitative} \cite{c9_dellermann2021finding}.
The majority of the work we reviewed \cite{c23_kim2017does,c20_lee2018content,c22_yu2018prediction,c3_sharchilev2018web,w3_gastaud2019varying,c21_cheng2019success,j46_kim2020recommendation,c6_ghassemi2020automated,j41_yeh2020machine,j42_srinivasan2020ensemble,j44_kaminski2020predicting,c9_dellermann2021finding,c15_ferrati2021deep,c19_garkavenko2021valuation,c16_chen2021trend,j2_ross2021capitalvx,c1_zhang2021scalable,j26_bai2021startup,j37_kinne2021predicting,j43_shi2021leveraging,t1_stahl2021leveraging,t2_horn2021deep,w2_yin2021solving,w1_lyu2021graph,j16_allu2022predicting,j9_tang2022deep,b1_ang2022using,j47_wu2022estimating,w6_garkavenko2022you} incorporate both modes into the model input, but they have to quantify the ``soft'' information via either approximation or questionnaire.
Data is often fed into DL model in the form of {\it feature}s.
Feature (a.k.a. ``factor'' in the scope of financial research) is an individual measurable property or characteristic of a phenomenon, which is sometimes aggregated from raw data.
When we try to map out the large number of features used in \cite{c23_kim2017does,c20_lee2018content,c22_yu2018prediction,c3_sharchilev2018web,w3_gastaud2019varying,c21_cheng2019success,j46_kim2020recommendation,c6_ghassemi2020automated,j41_yeh2020machine,j42_srinivasan2020ensemble,j44_kaminski2020predicting,c9_dellermann2021finding,c15_ferrati2021deep,c19_garkavenko2021valuation,c16_chen2021trend,j2_ross2021capitalvx,c1_zhang2021scalable,j26_bai2021startup,j37_kinne2021predicting,j43_shi2021leveraging,t1_stahl2021leveraging,t2_horn2021deep,w2_yin2021solving,w1_lyu2021graph,j16_allu2022predicting,j9_tang2022deep,b1_ang2022using,j47_wu2022estimating,w6_garkavenko2022you}, we found that features tend to cluster into different categories, describing different aspects of the startup in scope.  
We identified 15 natural categories (termed {\it funding}, {\it product/service}, {\it meta information}, {\it founder/owner}, {\it team}, {\it investor}, {\it web}, {\it context}, {\it connection}, {\it operation/planning}, {\it IP and R\&D}, {\it customer}, {\it financial}, {\it M\&A}\footnote{M\&A (merger and acquisition) refers to a business transaction in which the ownership of companies (or their operating units) are transferred to or consolidated with another company.} and {\it data}) and visualize their adoption percentage in Figure~\ref{fig:data-category-modality-dist}.
We hereby walk through each category following the order of popularity.

\subsection{A detailed walk-through of each data category}
\label{sec:walkthrough-data-category}
Historical {\bf funding} is direct evidence of recognition from other early investors, 
thus it is the most popular category in the literature.
Frequently seen features in this category are {\small \it total number of funding rounds and total amount raised} \cite{c3_sharchilev2018web,w3_gastaud2019varying,c9_dellermann2021finding,j2_ross2021capitalvx,t1_stahl2021leveraging,t2_horn2021deep,w1_lyu2021graph,w2_yin2021solving,j16_allu2022predicting,b1_ang2022using},
{\small \it funding types (e.g. angel, series A/B/C, debt financing, etc.)} \cite{c3_sharchilev2018web,c9_dellermann2021finding,c19_garkavenko2021valuation,j2_ross2021capitalvx,t1_stahl2021leveraging,w3_gastaud2019varying,j41_yeh2020machine,b1_ang2022using},
{\small \it elapsed time since latest funding} \cite{c3_sharchilev2018web,w3_gastaud2019varying,c19_garkavenko2021valuation,t1_stahl2021leveraging,b1_ang2022using,w6_garkavenko2022you},
{\small \it size and type of the latest funding} \cite{w3_gastaud2019varying,j2_ross2021capitalvx,b1_ang2022using,w6_garkavenko2022you},
{\small \it size and type of seed funding} \cite{c9_dellermann2021finding,j26_bai2021startup,w1_lyu2021graph},
{\small \it average per-round statistics} \cite{c19_garkavenko2021valuation,b1_ang2022using,w6_garkavenko2022you},
{\small \it average time between consecutive rounds} \cite{c3_sharchilev2018web,c19_garkavenko2021valuation,j2_ross2021capitalvx},
{\small \it the raw time-series of funding rounds} \cite{c16_chen2021trend,t1_stahl2021leveraging,t2_horn2021deep},
{\small \it accumulated amount for different funding types} \cite{c3_sharchilev2018web,j2_ross2021capitalvx},
{\small \it amount raised from VC} \cite{c9_dellermann2021finding,j2_ross2021capitalvx}, and
{\small \it the actual received amount of money} \cite{c3_sharchilev2018web,w6_garkavenko2022you}.
There are also some less common factors: 
{\small \it max/min round} \cite{w6_garkavenko2022you},
{\small \it raised amount in different currencies},
{\small \it information of the undisclosed rounds}, 
{\small \it elapsed time from seed funding} \cite{c3_sharchilev2018web},
{\small \it post-money valuation of rounds} \cite{c19_garkavenko2021valuation}, and
{\small \it alliance via M\&A} \cite{j2_ross2021capitalvx}.

\begin{figure}[t!]
\centering
\subcaptionbox{\footnotesize Distribution of the category of input data.\label{fig:data-category-dist}}
{\includegraphics[height=6cm]{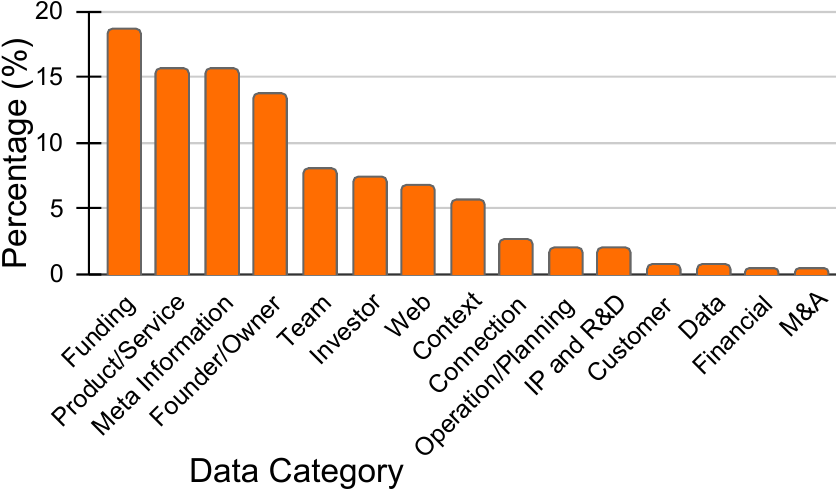}}
\hspace{1.5cm}
\subcaptionbox{\footnotesize Data Modality.\label{fig:data-modality-dist}}
{\includegraphics[height=6cm]{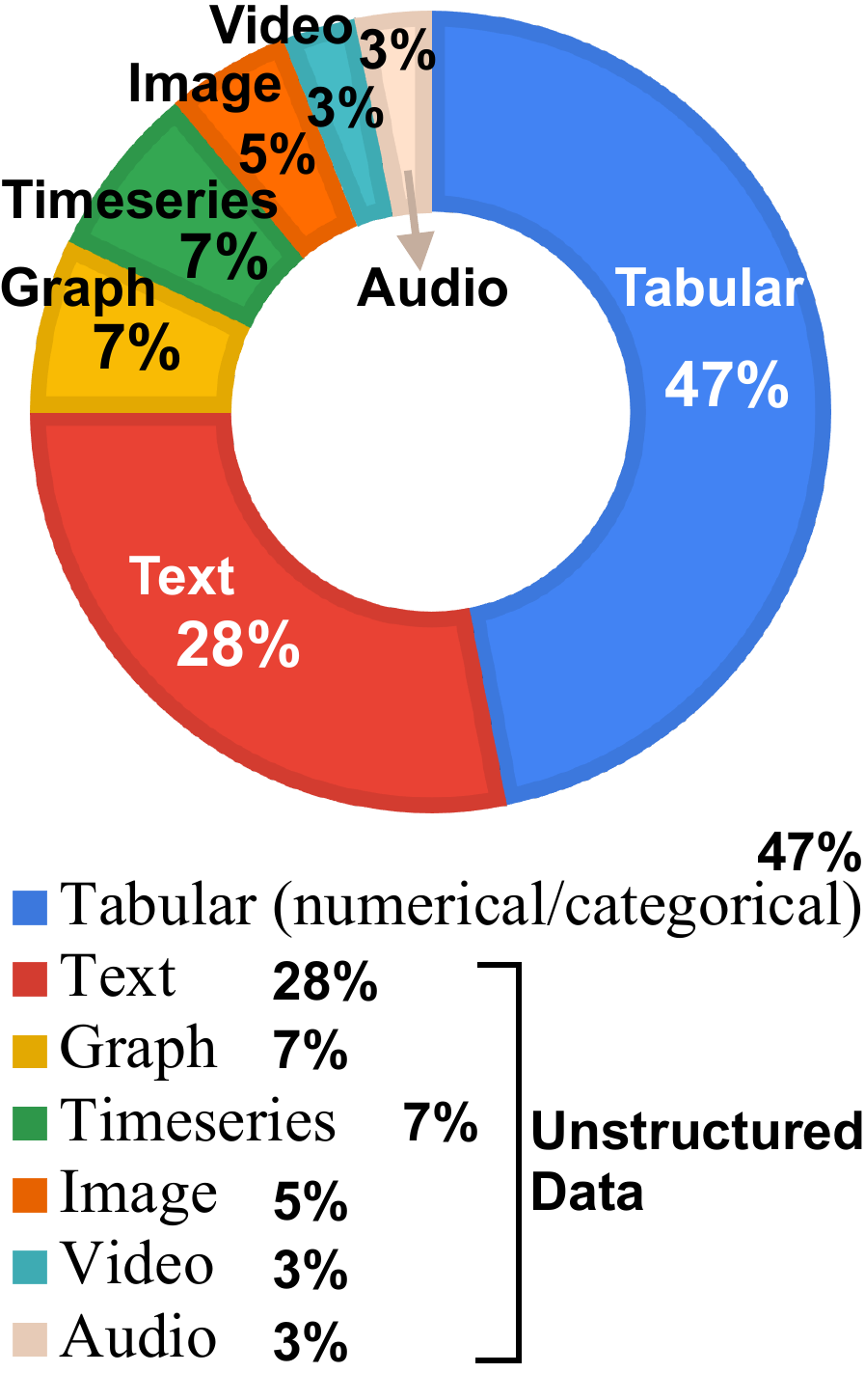}}
\caption{\label{fig:data-category-modality-dist}\textbf{Summary of the used categories of input data by surveyed work.}\newline (a) shows the percentage of each data category (detailed in Section~\ref{sec:walkthrough-data-category}) sorted by the their occurrences. (b) shows a snapshot (to the date when this paper is written) of the utilized data modalities: numerical, categorical, text, graph, time-series, image, video and audio.}
\end{figure}

The core value that early startups have to offer is reflected in the {\bf product/service} they aim to create, which makes this category of data widely adopted.
The top-3 features are {\small \it industry/sector/sub-sector} \cite{c3_sharchilev2018web,c6_ghassemi2020automated,c9_dellermann2021finding,c22_yu2018prediction,j2_ross2021capitalvx,j42_srinivasan2020ensemble,j43_shi2021leveraging,j47_wu2022estimating,t1_stahl2021leveraging,w1_lyu2021graph,b1_ang2022using},
{\small \it textual description} \cite{c6_ghassemi2020automated,c16_chen2021trend,c20_lee2018content,c21_cheng2019success,c23_kim2017does,j37_kinne2021predicting,j44_kaminski2020predicting,j46_kim2020recommendation,j47_wu2022estimating}, and
{\small \it project specification on crowdfunding platforms} \cite{c21_cheng2019success,c22_yu2018prediction,c23_kim2017does,j41_yeh2020machine,j42_srinivasan2020ensemble,j43_shi2021leveraging,j47_wu2022estimating}. 
DL models are also suited well for learning representations from {\small \it image, video and audio} \cite{c21_cheng2019success,j9_tang2022deep,j42_srinivasan2020ensemble,j43_shi2021leveraging,j44_kaminski2020predicting}.
The rest of the features describing product/service include {\small \it technology maturity} \cite{c9_dellermann2021finding,j16_allu2022predicting},
{\small \it customer focus (e.g. B2B/B2C/B2B2C)} \cite{c9_dellermann2021finding,t1_stahl2021leveraging},
{\small \it time to market} \cite{c9_dellermann2021finding,c3_sharchilev2018web},
{\small \it novelty and differentiation} \cite{c9_dellermann2021finding,j26_bai2021startup},
{\small \it quality measure (e.g. simplicity and usability)},
{\small \it market penetration and traction} \cite{j26_bai2021startup},
{\small \it business scalability},
{\small \it business models (e.g. subscription centric, freemium, cross selling, hidden revenue, no frills, layer player)} \cite{c9_dellermann2021finding},
{\small \it the number of product varieties} \cite{c3_sharchilev2018web},
{\small \it textual product review and comment} \cite{c20_lee2018content}, and
{\small \it idea rating by experts} \cite{c6_ghassemi2020automated}.

{\bf Meta information} refers to the general attributes about startups, which seldomly changes since creation/registration of the firm.
Most work use factors of {\small \it elapsed time since founded}, {\small \it textual description}, and {\small \it geographical location} \cite{c3_sharchilev2018web,c9_dellermann2021finding,c16_chen2021trend,c19_garkavenko2021valuation,c22_yu2018prediction,j2_ross2021capitalvx,j9_tang2022deep,j42_srinivasan2020ensemble,j43_shi2021leveraging,t1_stahl2021leveraging,t2_horn2021deep,w1_lyu2021graph,w2_yin2021solving,w3_gastaud2019varying,w6_garkavenko2022you,b1_ang2022using}.
Whether a startup {\small \it has Facebook/Linkedin/Twitter account} \cite{c9_dellermann2021finding,c23_kim2017does,j2_ross2021capitalvx,j43_shi2021leveraging,w6_garkavenko2022you} is also a common factor.
Other less frequently seen factors include {\small \it domain name or homepage URL} \cite{c23_kim2017does,j2_ross2021capitalvx,j42_srinivasan2020ensemble},
{\small \it company name and aliases} \cite{j2_ross2021capitalvx,j42_srinivasan2020ensemble},
{\small \it office count and age} \cite{c3_sharchilev2018web,w6_garkavenko2022you},
{\small \it registered address},
{\small \it current status (e.g. operating, closed, zombie)},
{\small \it official email and phone number} \cite{j2_ross2021capitalvx}, and
{\small \it incubator or accelerator support} \cite{c9_dellermann2021finding}.

The traits of a startup's {\bf founder/owner} (i.e. the entrepreneur or the founding team) are so important that the founder(s) with relevant management experiences can improve the company performance \cite{b2_corea2019ai,ewens2018founder}.
The attributes of founding teams and the individuals that comprise them contribute to their short-term success and longer-term survival \cite{c6_ghassemi2020automated}, and are also generally available from many data sources and entrepreneurial competitions.
The {\small \it founding team size (number of co-founders)} and {\small \it founders' (successful) founding/industry experience} \cite{c3_sharchilev2018web,c9_dellermann2021finding,c19_garkavenko2021valuation,j2_ross2021capitalvx,j26_bai2021startup,j41_yeh2020machine,j42_srinivasan2020ensemble,j43_shi2021leveraging,j46_kim2020recommendation,w2_yin2021solving,w3_gastaud2019varying} are most widely used, followed by 
{\small \it founder IDs from data sources} \cite{c3_sharchilev2018web,j41_yeh2020machine,j42_srinivasan2020ensemble}, 
{\small \it gender/ethnicity} \cite{c3_sharchilev2018web,j2_ross2021capitalvx,w1_lyu2021graph,j30_kaiser2020value}, and
{\small \it social capital} \footnote{Social capital is a positive product of human interactions, which comprises two aspects: bonding (intra group) and bridging (inter groups). Nowadays, it is increasingly represented by activities on social media and applications \cite{j43_shi2021leveraging}.} \cite{j42_srinivasan2020ensemble,j43_shi2021leveraging}.
Additionally, some researchers \cite{c6_ghassemi2020automated,j26_bai2021startup,t5_bento2018predicting,rj1_pasayat2020factors} try to quantify the {\small \it founders' skill (e.g. leadership, research, development, product management, sales, law, consulting, finance, marketing, creativity and investment)}, and use it as model input.
The rest of the factors in this category include 
{\small \it years between graduation and founding},
{\small \it education institute and major} \cite{c6_ghassemi2020automated,b2_corea2019ai,t5_bento2018predicting},
{\small \it founders' biography and photo} \cite{j42_srinivasan2020ensemble,c23_kim2017does}, and finally
indications of founders' {\small \it entrepreneurial vision} \cite{c9_dellermann2021finding}, {\small \it capability of work (dedication)} \cite{j26_bai2021startup}, and {\small \it 3rd-party score} \cite{c3_sharchilev2018web,j43_shi2021leveraging}.

Complementary to founder data, {\bf team} related factors are used in many research papers.
The common factors are {\small \it team size of all or different functions} \cite{c9_dellermann2021finding,c19_garkavenko2021valuation,j2_ross2021capitalvx,j46_kim2020recommendation,w6_garkavenko2022you,b1_ang2022using},
{\small \it the time-series of team size}
\cite{t1_stahl2021leveraging,t2_horn2021deep},
{\small \it statistics of new hire or leavers} \cite{c3_sharchilev2018web,c19_garkavenko2021valuation}, 
{\small \it completeness and capability of managers} \cite{c19_garkavenko2021valuation,j26_bai2021startup}, and
{\small \it team composition (e.g. diversity and gender)} \cite{c3_sharchilev2018web,j2_ross2021capitalvx}
The less common ones are
{\small \it time of involvement},
{\small \it board member statistics},
{\small \it person IDs from data sources} \cite{c3_sharchilev2018web},
{\small \it vocational skill and experience} \cite{c19_garkavenko2021valuation},
{\small \it technical team size and quality},
{\small \it employees from renowned organizations} \cite{c16_chen2021trend},
{\small \it educational degrees of employees} \cite{j2_ross2021capitalvx},
{\small \it team constellation} \cite{c9_dellermann2021finding},
{\small \it balance/empowerment/competence of project team} \cite{j41_yeh2020machine}, and
{\small \it 3rd-party team score} \cite{c6_ghassemi2020automated}.

Closely related to funding data, statistics of the existing {\bf investor}(s) provide useful information about the startup.
We observe that {\small \it number of total/distinct investors} \cite{c3_sharchilev2018web,c15_ferrati2021deep,c16_chen2021trend,c22_yu2018prediction,j2_ross2021capitalvx,j46_kim2020recommendation,w3_gastaud2019varying,b1_ang2022using} is the most heavily used factor in this category.
Authors of \cite{c3_sharchilev2018web,c15_ferrati2021deep,t1_stahl2021leveraging,w2_yin2021solving} try to {\small \it rank investors by their reputation, experience, IPO/M\&A performance}.
Moreover, the concept of {\small \it VC syndicate (e.g. advantage, diversity, centrality and foreignness of VC)} is also investigated in \cite{j19_hochberg2007whom,j15_shin2019network,w3_gastaud2019varying,j34_nahata2008venture}.
In DL-based work, {\small \it share and involvement time of each investor} \cite{c3_sharchilev2018web} is also used.

{\bf Web} category embodies information of web pages about the startup. The literature has mentioned several factors: {\small \it rank/count/duration/bounce rate (aggregated or time-series) of website visit} \cite{c3_sharchilev2018web,c9_dellermann2021finding,t1_stahl2021leveraging,t2_horn2021deep,w6_garkavenko2022you},
{\small \it news count} \cite{c3_sharchilev2018web,c19_garkavenko2021valuation,w2_yin2021solving,w3_gastaud2019varying},
{\small \it topics of news/articles} \cite{c3_sharchilev2018web,j46_kim2020recommendation,w6_garkavenko2022you},
{\small \it Twitter statistics (e.g. followers, tweets and sentiment)} \cite{c9_dellermann2021finding,c19_garkavenko2021valuation,w6_garkavenko2022you}, and
{\small \it count of websites, pages and domain names that mentioned the firm} \cite{c3_sharchilev2018web,c9_dellermann2021finding,w6_garkavenko2022you}.

\begin{figure}[t!]
\centering
\subcaptionbox{\footnotesize Connection among entities. \label{fig:connection-illustration}}
{\includegraphics[height=4.5cm]{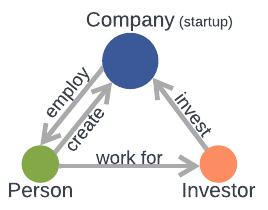}}
\hfill
\subcaptionbox{\footnotesize An example of company-person-investor graph.\label{fig:connection-example}}
{\includegraphics[height=4.5cm]{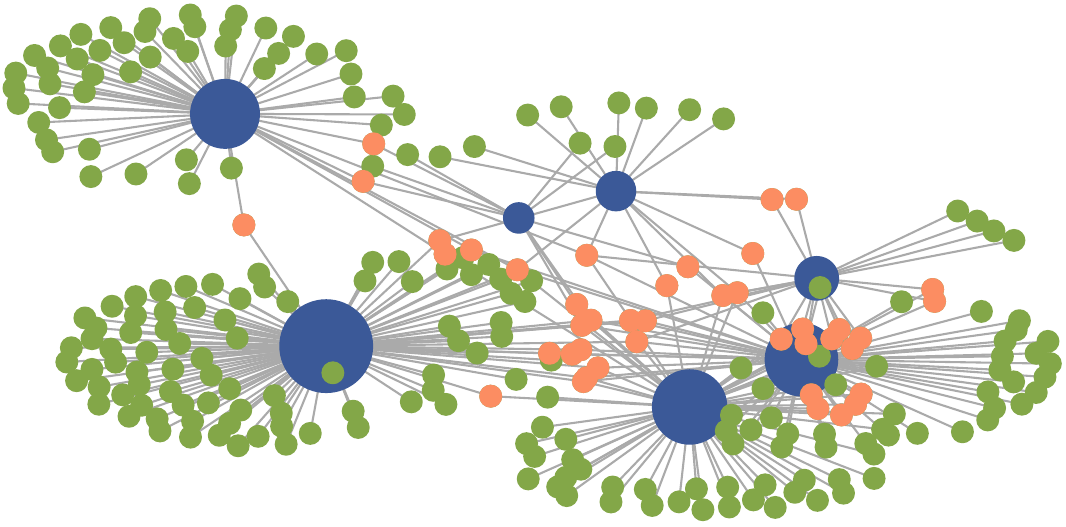}}
\caption{\label{fig:connection-illustration-example}\textbf{Illustration of connection data category where a graph is can be constructed.}\newline The graph comprises nodes (denoting company/person/investor) and edges (representing investing/employment/founding relations between nodes). To understand the example graph in (b), refer to the color coding in (a) as a legend.}
\end{figure}

So far, we have only touched upon the factors that are intrinsic to the startup in consideration, but more and more researchers have realized the importance of extrinsic factors\footnote{While intrinsic factors act from within a company, extrinsic factors wield their influence from the outside. The former can be controlled (to some extent) by the startup-in-scope, but the latter can not, since they represent external contexts that may be (but not limited to) competition, environmental, cultural, economical and tax-based.}. In this paper, we put each extrinsic factor into one of the two categories: {\bf context} and {\bf connection}.
The top-3 {\bf context} features are {\small \it the number of direct competitors} \cite{c2_pasayat2021evolutionary,c3_sharchilev2018web,c5_xiang2012supervised,c9_dellermann2021finding,j16_allu2022predicting,j26_bai2021startup,j46_kim2020recommendation,w3_gastaud2019varying},
{\small \it funding raised by competitors} \cite{t1_stahl2021leveraging,w3_gastaud2019varying}, and
{\small \it per-industry prosperity of the hosting geo-location} \cite{w2_yin2021solving,w3_gastaud2019varying}.
Besides, there are other context factors, such as 
{\small \it market/industry size and growth rate},
{\small \it exchange rate},
{\small \it inflation level},
{\small \it governmental regulation},
{\small \it tax policy} \cite{j16_allu2022predicting},
{\small \it sector performance},
{\small \it country/state economy} \cite{j2_ross2021capitalvx},
{\small \it financing environment} \cite{w2_yin2021solving}, and
{\small \it current month/week} \cite{t2_horn2021deep}.
The {\bf connection} features are usually extracted from a graph (cf. Figure~\ref{fig:connection-example}) that encodes connections between different entities: startup, person and investor, as illustrated in Figure~\ref{fig:connection-illustration}. 
The DL approaches \cite{c1_zhang2021scalable,c16_chen2021trend,w1_lyu2021graph,w3_gastaud2019varying}
often directly feed the graph into ANNs, while ML-based methods (e.g. \cite{j27_liang2016predicting,j17_bonaventura2020predicting,j19_hochberg2007whom}) always pre-calculate some graph features, such as betweenness/closeness/degree centrality, shortest paths, common neighbors, etc.

{\bf Operation/planning} typically involves operational matters such as  sales, localization (e.g. \cite{c2_pasayat2021evolutionary,rj1_pasayat2020factors}), marketing (e.g. \cite{j26_bai2021startup,j33_malmstrom2020they}), supply chain \cite{rj1_pasayat2020factors,rj2_song2008success}, digitization \cite{j36_rouhani2013erp}, advisory \cite{c5_xiang2012supervised}, company culture (e.g. \cite{j8_santisteban2021critical,w5_guerzoni2019survival}) and legal regulation \cite{j36_rouhani2013erp,j22_skawinska2020success}. 
However, the DL-based methods have only a few mentions of factors in this category:
{\small \it planned revenue model} \cite{c9_dellermann2021finding,j16_allu2022predicting,j26_bai2021startup},
{\small \it global exposure and internationalization} \cite{c3_sharchilev2018web},
{\small \it market positioning and go-to-market strategy} \cite{j26_bai2021startup}, and
{\small \it technological surveillance} \cite{j16_allu2022predicting}.
For early startups, IP (intellectual property) and R\&D (research and development) are two of the aspects that are examined by investors.
As a result, we make {\bf IP and R\&D} its own category (out of the operational factors), which contains {\small \it number of patents}, {\small \it patent growth}, {\small \it patent category} \cite{c15_ferrati2021deep,j2_ross2021capitalvx,j37_kinne2021predicting,j46_kim2020recommendation} and
{\small \it university partnership} \cite{c9_dellermann2021finding}.

The {\bf customer}, {\bf financial} and {\bf M\&A} data is, most of the time, unavailable publicly,
which resonates with their scarce occurrence in Figure~\ref{fig:data-category-dist}.
For {\bf customer} category, we have seen {\small \it customer satisfaction/loyalty} \cite{c16_chen2021trend} and {\small \it the number of pilot customers} \cite{c9_dellermann2021finding}.
The most common {\bf financial} factor turns out to be {\small \it revenue and turnover}~\cite{j46_kim2020recommendation,cao-etal-2022-sire}.
In {\bf M\&A} category, \cite{j2_ross2021capitalvx} calculated the {\small \it the number of acquisitions} as an input feature to their model.
Finally, the {\bf data} category contains statistics that are specific to the chosen data source, such as {\small \it the total number of events/records} \cite{j46_kim2020recommendation} in Crunchbase.

\subsection{Several noticeable trends in data selection}
The surveyed literature reflects several trend concerning selecting the input data for DL models.
We summarize these trends as a guidance for investment practitioners.

\begin{itemize}
\item {\bf Single-modal$\rightarrow$multi-modal}\footnote{Modality refers to the way in which data is generated or presented, and a research is characterized as multi-modal when it includes multiple modalities such as text and image.}: although the {\small \it tabular (aggregated numerical/categorical data)} form still dominates, we see other emerging data modalities: {\small \it text}, {\small \it graph}, {\small \it time-series}, {\small \it image}, {\small \it video} and {\small \it audio}. The relative adoption of different modalities is shown in Figure~\ref{fig:data-modality-dist}. Especially, a few recent work \cite{j43_shi2021leveraging,j44_kaminski2020predicting,c21_cheng2019success,t1_stahl2021leveraging,t2_horn2021deep} has looked into combining multiple input modalities (i.e. multi-modal).

\item {\bf Structured(aggregated)$\rightarrow$unstructured(raw)}\footnote{Unstructured data, such as image and timeseries, is a collection of many varied types that maintains their native and original form, while
structured data is aggregated from original (raw) data and is usually stored in a tabular form.}: 
the modalities excluding ``tabular'' in Figure~\ref{fig:data-modality-dist} are all unstructured, which become increasingly important as a complement to the structured data (e.g. \cite{t1_stahl2021leveraging,t2_horn2021deep,w1_lyu2021graph,w3_gastaud2019varying,c16_chen2021trend}), or as a standalone input to the model (e.g. \cite{c1_zhang2021scalable,j9_tang2022deep}).
Since raw, unstructured data often has a large scale and contains intact-yet-noisy signal, it may bring forward superior performance as long as a proper DL approach is applied \cite{w6_garkavenko2022you}.

\begin{figure}[t!]
\centering
\subcaptionbox{\footnotesize Distribution of used data sources (sorted).\label{fig:datasource-dist}}
{\includegraphics[height=6.2cm]{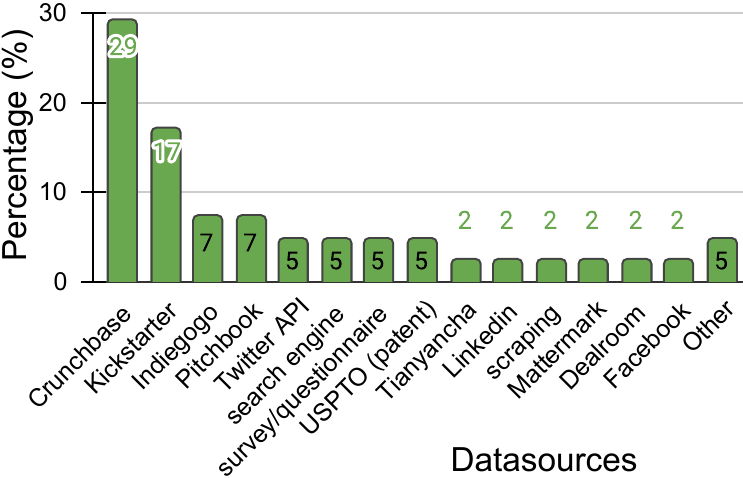}}
\hfill
\subcaptionbox{\footnotesize Dataset size distribution\label{fig:dataset-scale}}
{\includegraphics[height=6.2cm]{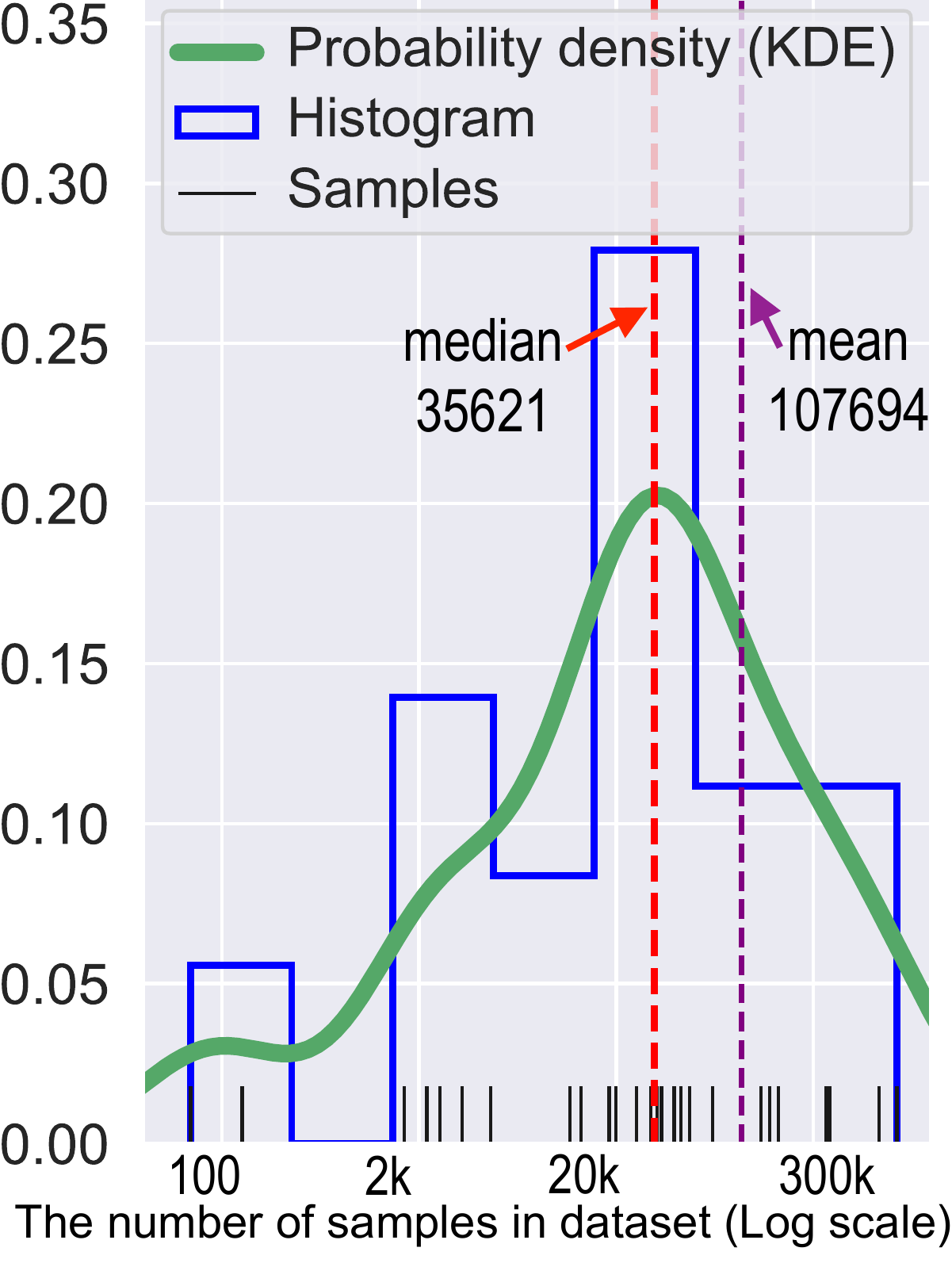}}
\caption{\label{fig:dataset-dist-scale}\textbf{Statistics about datasets: (a) occurrences and (b) the number of samples.}\newline In (a), {\bf paid} data sources are Crunchbase (\protect\url{crunchbase.com}), Pitchbook (\protect\url{pitchbook.com}), Tianyacha (\protect\url{tianyancha.com}), Linkedin (\protect\url{linkedin.com}), Mattermark (\protect\url{mattermark.com}), Dealroom (\protect\url{dealroom.co});
{\bf free} sources are Kickstarter/Indiegogo/scraping (\protect\url{webrobots.io}), Twitter API (\protect\url{developer.twitter.com}), search engines (e.g. \protect\url{google.com}), USPTO (United States Patent and Trademark Office: \protect\url{uspto.gov}), Facebook (only the pages about startups); 
{\bf proprietary} data are usually only accessible from investment firms (in ``Other'' category), governmental/administrative departments or survey/questionnaire.
In (b), we plot the distribution (probability density), histogram, median, mean, maximum (776,273) and minimum (100) of dataset size (i.e. the number of records).}
\end{figure}

\item {\bf Proprietary$\rightarrow$paid$\rightarrow$free}: all data sources utilized in DL-based methods are sorted in Figure~\ref{fig:datasource-dist} according to their occurrences. 
The traditional proprietary sources are not favored any more due to the limitation of scale and shareability. 
Paid data sources (e.g. Crunchbase and Pitchbook) are still very popular, because they are mostly quite affordable and well organized. 
However, neither paid or proprietary data is up-to-date or fine-grained, which contributes to the increasing adoption of free data sources, such as web page scraping \cite{w6_garkavenko2022you}.

\item {\bf Gather at least 35k samples}: to understand how many samples (often correlated with the number of companies) researchers use for training their DL models, we plot the distribution/histogram of dataset size in Figure~\ref{fig:dataset-scale}. It shows a median and average size of 35,621 and 107,694 respectively. It has been discussed previously that DL model generally require more data to match with its big capacity$^7$, thus we recommend to collect at least 35k samples. 

\item {\bf Intrinsic(independent)$\rightarrow$extrinsic(contextual)}: classically, most elements driving investors’ decisions would seem to be only {\it independent and intrinsic}$^{18}$ to the startup in scope, most notably at the expense of {\it extrinsic and contextualized}$^{18}$ features \cite{w3_gastaud2019varying}. 
The community has realized this, 
and steers towards using more {\bf context} and {\bf connection} (cf. Figure~\ref{fig:connection-illustration-example}) data to complement the intrinsic features.
\end{itemize}

\section{Address the Problems of Data Imbalance and Sparsity}
During the process of preparing the training data for DL models, one is almost certain to encounter two problems: data {\it imbalance} and {\it sparsity}.
Therefore we will discuss the causes of these two problems and present some effective solutions.

\begin{figure}[!t]
\centerline{\includegraphics[width=\linewidth]{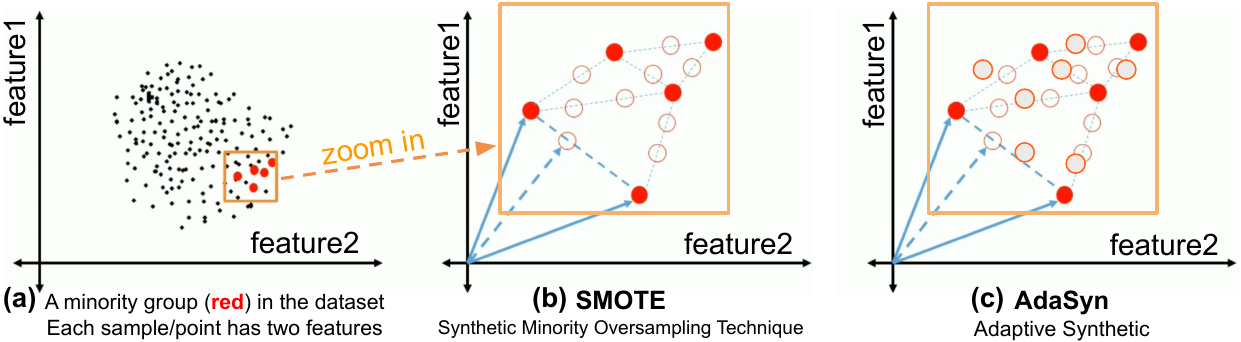}}
\caption{\label{fig:data-imbalance}\textbf{Re-balance the dataset by augmenting the minority samples.}\newline For the sake of visualization, we assume the dataset only has two dimensions (feature 1 and 2). The minority group (e.g. success startups) is indicated with red color in (a). SMOTE randomly sample new points {\bf on} the lines between existing minority samples, and AdaSyn allows sampling of new points {\bf off} those lines from a Gaussian distribution.}
\end{figure}

\subsection{Balance the dataset with augmentation or PU-learning}
The purpose of clearly defining the startup success criteria (Section~\ref{sec:success_criteria}) is to assign a label (either positive/successful or negative/unsuccessful) for each sample/record in the dataset.
Since successful startups are inherently rare$^2$, the number of positive samples [e.g. the red points in Figure~\ref{fig:data-imbalance}(a)] is significantly less than the negative ones.
Another common formality of data imbalance concerns the drastically different population size in different groups (by sector, customer focus, country, etc.), for example, a dataset which contains mostly fintech companies mixed with only a few biotech companies [cf. the red points in Figure~\ref{fig:data-imbalance}(a)].
Without special treatment, DL models will exhibit bias towards the majority class, and in extreme cases, may ignore the minority class altogether \cite{johnson2019survey}.

\begin{figure}[!t]
\centerline{\includegraphics[width=\linewidth]{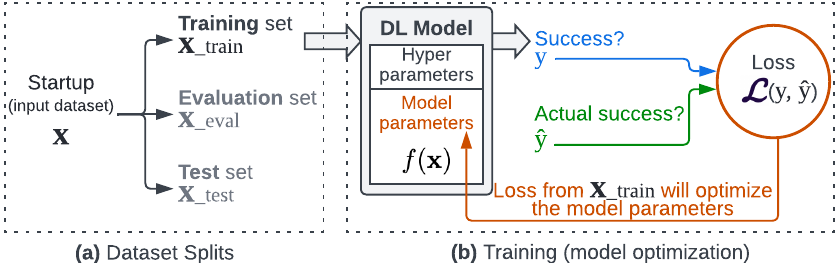}}
\vspace{5pt}
\caption{\label{fig:data-split-train-process}\textbf{High-level illustration of (a) dataset splitting and (b) training procedure.}\newline The entire dataset $\mathbf{x}$ should be randomly divided into three splits: $\mathbf{x}_\text{train}$, $\mathbf{x}_\text{eval}$ and $\mathbf{x}_\text{test}$. The training subset $\mathbf{x}_\text{train}$ is used for training the DL model, where a loss calculated to measure the prediction error (the deviation from the ground-truth label).
The loss will guide the optimization of the model parameters.} 
\vspace{10pt}
\end{figure}

The literature \cite{w2_yin2021solving,j2_ross2021capitalvx,t3_Unal2019Machine,t5_bento2018predicting,t6_unal2019searching,w5_guerzoni2019survival,c15_ferrati2021deep,t3_Unal2019Machine,t8_kamal2021modeling,w6_garkavenko2022you} mentioned three approaches to ``rebalance'' an imbalanced dataset.
\begin{itemize}
\item {\bf SMOTE} (synthetic minority oversampling technique) \cite{chawla2002smote,w2_yin2021solving,j2_ross2021capitalvx,t3_Unal2019Machine,t5_bento2018predicting,t6_unal2019searching,w5_guerzoni2019survival}: As demonstrated in Figure~\ref{fig:data-imbalance}(b), random samples are drawn along the lines connecting the pairs of minority samples.
\item {\bf AdaSyn} (adaptive synthetic) \cite{he2008adasyn,c15_ferrati2021deep,t3_Unal2019Machine,t6_unal2019searching,t8_kamal2021modeling}: Different from SMOTE, AdaSyn allows the newly sampled points [the red circles filled with gray color in Figure~\ref{fig:data-imbalance}(c)] to deviate (obeying a Gaussian distribution) from the lines between existing minority samples.
\item {\bf PU-learning} (positive unlabeled learning) \cite{kiryo2017positive,w6_garkavenko2022you}: Instead of generating additional synthetic samples, PU-learning modifies the loss function [cf. Figure~\ref{fig:data-split-train-process}(b)] so that it 1) allows unlabeled samples to participate in the training, and 2) focuses more on the minority samples throughout the optimization.
\end{itemize}

It is worth mentioning that SMOTE and AdaSyn can merely augment tabular (numeric) features. 
More advanced techniques are required to augment unstructured raw data (e.g. text and images).
For instance, text can be augmented with synonym replacement, insertion, swap, deletion \cite{wei_zou_2019_eda}, and summarization \cite{yao2017recent}.

\subsection{Densify sparse input with simple imputation techniques}
\begin{figure}[!t]
\centerline{\includegraphics[width=0.8\linewidth]{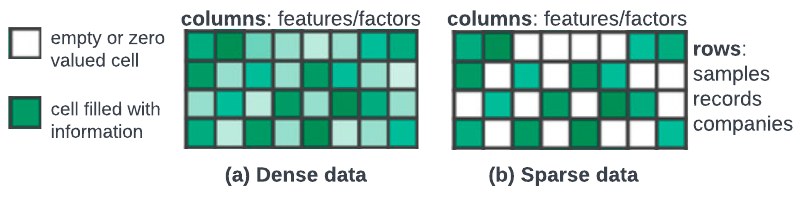}}
\vspace{-3pt}
\caption{\label{fig:dense-vs-sparse}\textbf{An illustration of dense data vs. sparse data.}\newline A row denotes an individual sample, data record or company, while a column represents a certain feature or factor. A white-colored cell indicates a missing or zero value.}
\end{figure}

Sparsity and density describe the amount of features/factors (columns in Figure~\ref{fig:dense-vs-sparse}) in a dataset that are ``empty'' (sparse) and filled with information (dense), though ``empty'' could also mean zero-valued.
Since early-stage startup companies do not have much data available to the public, the resulting dataset (for startup success prediction) is typically sparse as exemplified in Figure~\ref{fig:dense-vs-sparse}(b).
While some ML algorithms are sparse-aware (e.g. XGBoost) \cite{w2_yin2021solving}, DL models (ANNs) rely on spatial or sequential correlations to learn and highly sparse input will break these correlations. As a result, ANNs are not sparsity agnostic.

A straightforward way to avoid sparse input is to remove the entire sample/record (row in Figure~\ref{fig:dense-vs-sparse}) or feature/factor (column in Figure~\ref{fig:dense-vs-sparse}) that contain ``empty'' cell(s).
But a drastically reduced dataset scale prohibits its adoption in reality.
To match the dataset scale with model capacity, we need to keep as many rows and/or columns (Figure~\ref{fig:dense-vs-sparse}) as possible. 
Imputation is widely utilized to achieve that goal.
Although there exist complex imputation methods based on DL models like autoencoders, there is no guarantee (e.g. \cite{jager2021benchmark}) that they outperform simpler ones.
Instead, literature suggests to use simple methods, such as {\it mean/mode/zero imputation} \cite{t1_stahl2021leveraging,t5_bento2018predicting}, {\it latest-value imputation} \cite{t2_horn2021deep}, {\it Soft-Impute} \cite{mazumder2010spectral,t7_saini2018picking}, and {\it $k$-NN ($k$ nearest neighbor) imputation} \cite{w2_yin2021solving}.

\section{Split the Dataset with an Investor-Centric View}
\label{sec:dataset-split}
Splitting the dataset is a mandatory step before training any ML/DL model, yet it is often discussed very lightly (sometimes even neglected) in the literature on startup success prediction.
It is generally recommended to divide the dataset into non-overlapping {\it training} ($\mathbf{x}_\text{train}$), {\it evaluation} ($\mathbf{x}_\text{eval}$) and {\it test} ($\mathbf{x}_\text{test}$) subsets, as shown in Figure~\ref{fig:data-split-train-process}(a).
The model will be trained solely on the training set, during which the {\it model parameters} will be optimized.
But as illustrated in the left part of Figure~\ref{fig:data-split-train-process}(b), the DL model also has {\it hyper-parameters}\footnote{Hyper-parameters are parameters controlling the learning process, hence also indirectly determining the values of model parameters. After the model is trained, hyper-parameters will not participate the model inference, thus they are not a part of the model parameters.} to be tuned in a process called hyper-parameter search. 
In the simplest form, the training will be run for $N$ times with different hyper-parameters, resulting in $N$ trained models, each of which is evaluated on $\mathbf{x}_\text{eval}$.
The best performing model on $\mathbf{x}_\text{eval}$ should be used for further testing and production.
The test set $\mathbf{x}_\text{test}$ should not be exposed to the hyper-parameter tuning process; it is used to report the performance of the chosen model.
Section~\ref{sec:eval} (Figure~\ref{fig:eval-test-process}) can be referred to for more details of the evaluation and testing processes.

\subsection{Company-centric vs. investor-centric}
\begin{figure}[!t]
\centerline{\includegraphics[width=\linewidth]{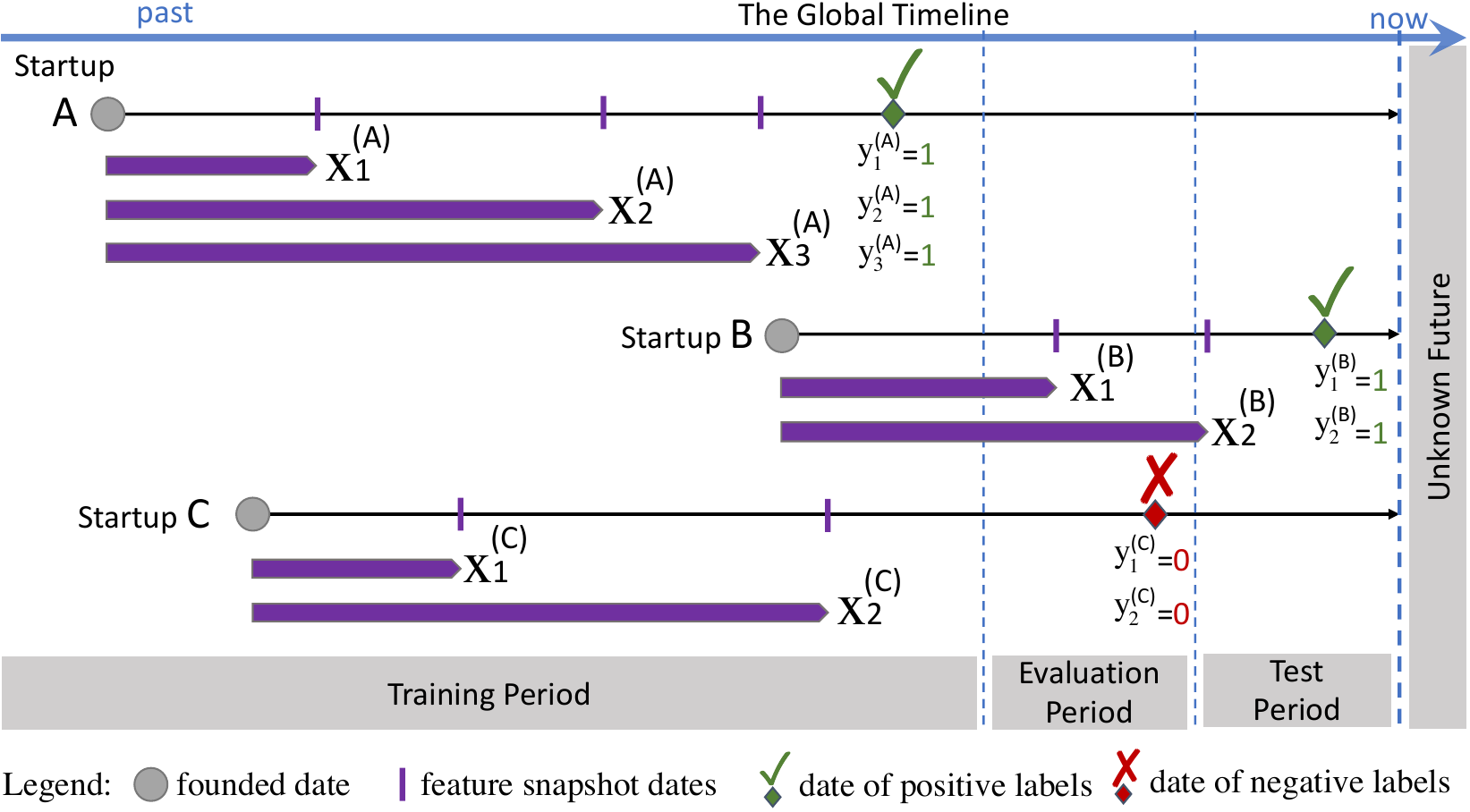}}
\vspace{-3pt}
\caption{\label{fig:data-split-demo}\textbf{Visualization of (investor-centric) dataset split using three example startups.}\newline A startup (A/B/C) obtains a positive/negative label on a certain date, before which several feature snapshot dates are picked.
For each snapshot date, a sample is computed using data available up till that date.
In this manner, A generates three $\langle$sample-label$\rangle$ pairs: $\langle\mathbf{x}_1^{(\text{A})}, y_1^{(\text{A})}\rangle$, 
$\langle\mathbf{x}_2^{(\text{A})}, y_2^{(\text{A})}\rangle$ and
$\langle\mathbf{x}_3^{(\text{A})}, y_3^{(\text{A})}\rangle$; 
B produces two pairs: 
$\langle\mathbf{x}_1^{(\text{B})}, y_1^{(\text{B})}\rangle$ and
$\langle\mathbf{x}_2^{(\text{B})}, y_2^{(\text{B})}\rangle$;
C also spawns two pairs:
$\langle\mathbf{x}_1^{(\text{C})}, y_1^{(\text{C})}\rangle$ and
$\langle\mathbf{x}_2^{(\text{C})}, y_2^{(\text{C})}\rangle$.
The global timeline is fragmented into training, evaluation and test periods. 
The period that the label (in the $\langle$sample-label$\rangle$ pairs) belongs determines the particular training/evaluation/test set each pair should go to.} 
\end{figure}

To predict the success of startups, the appropriate way to split the dataset is not as straightforward as it appears in ML/DL researches for other domains.
We visualize a minimal example in Figure~\ref{fig:data-split-demo} to facilitate our discussion;
there are three startups (A, B and C) founded at different dates over the timeline. 
According to some predefined success criteria (Section~\ref{sec:success_criteria}), A and B are labeled as positive (i.e. promising investing targets: $y^{(\text{A})}=y^{(\text{B})}=1$) some time after they are founded. 
The majority become unfavourable (e.g. the label of C is $y^{(\text{C})}=0$) to VC firms, if no sign of success some years after their founding dates. 

With a {\bf company-centric view}, one could choose some event types (e.g. seed and pre-A rounds), the dates of which are called {\it feature snapshot dates}.
We can then compute one sample using data before each snapshot date.
As shown in Figure~\ref{fig:data-split-demo}, there are three snapshot dates on the timeline of startup A, leading to three samples (i.e. $\mathbf{x}_1^{(\text{A})}$, $\mathbf{x}_2^{(\text{A})}$ and $\mathbf{x}_3^{(\text{A})}$) that are all labeled positive (i.e. $y_1^{(\text{A})}=y_2^{(\text{A})}=y_3^{(\text{A})}=1$). 
In a sense, A is augmented by generating three $\langle$sample-label$\rangle$ pairs:
$\langle\mathbf{x}_1^{(\text{A})}, y_1^{(\text{A})}\rangle$, 
$\langle\mathbf{x}_2^{(\text{A})}, y_2^{(\text{A})}\rangle$ and
$\langle\mathbf{x}_3^{(\text{A})}, y_3^{(\text{A})}\rangle$.
Similarly, B and C produce another four pairs: 
$\langle\mathbf{x}_1^{(\text{B})}, y_1^{(\text{B})}\rangle$,
$\langle\mathbf{x}_2^{(\text{B})}, y_2^{(\text{B})}\rangle$,
$\langle\mathbf{x}_1^{(\text{C})}, y_1^{(\text{C})}\rangle$ and
$\langle\mathbf{x}_2^{(\text{C})}, y_2^{(\text{C})}\rangle$.
The company-centric split \cite{b1_ang2022using,c19_garkavenko2021valuation,c20_lee2018content,c22_yu2018prediction,c9_dellermann2021finding,j41_yeh2020machine,j42_srinivasan2020ensemble,j43_shi2021leveraging,j44_kaminski2020predicting,j46_kim2020recommendation,w3_gastaud2019varying} will randomly allocate those pairs into one of the sets (training, evaluation or test).

With an {\bf investor-centric view} \cite{c15_ferrati2021deep,c16_chen2021trend,c21_cheng2019success,j47_wu2022estimating,w1_lyu2021graph,w2_yin2021solving,w6_garkavenko2022you}, the feature snapshot dates are randomly sampled (before the corresponding label date), therefore they do not represent any event(s).
More importantly, the global timeline is fragmented (from earliest startup founding date to now) into three periods, i.e. training, evaluation and test period, as illustrated in Figure~\ref{fig:data-split-demo}.
For a startup, the period that its label belongs determines the dataset split it should go to.
Using this rule, we can see (cf. Figure~\ref{fig:data-split-demo}) that the three $\langle$sample-label$\rangle$ pairs from A should go to the training set; the two pairs from B belong to the test set; and lastly, the two pairs from C will head to the evaluation set.
The {\bf investor-centric view is generally preferred}, because it better resembles the real-world scenario of how investment professionals predict the success of startup candidates \cite{c3_sharchilev2018web}.

\subsection{Understand the data generation process}
When assembling the samples (i.e. $\mathbf{x}_i^{(\cdot)}$ in Figure~\ref{fig:data-split-demo}) using data up till the snapshot dates, one should make sure that no future information is leaked into $\mathbf{x}_i^{(\cdot)}$.
This requires in-depth understanding of not only the data itself ({\it know-what}) but also the data generation process ({\it know-how}), which we found is seldomly addressed by the literature.
We hereby give a concrete example out of many: a startup in the dataset has an annual revenue data point (from BvD\footnote{Bureau van Dijk: \url{www.bvdinfo.com}}) with a timestamp 2020-12-31; but this data point should be ignored when predicting on 2021-06-01. 
The reason is that fiscal reports (the source of revenue data) often have a delay of more than 12 months, causing the 2020-12-31 data point unavailable until (earliest) 2021-12-31.
Without carefully examining the risk of information leakage of features/factors from the perspective of data generation process, the model performance in production may fail catastrophically.

\section{Model Selection: Occam's Razor and No-Free-Lunch}
\label{sec:model-selection}

\begin{figure}[t!]
\centering
\subcaptionbox{\footnotesize Distribution of DL model category.\label{fig:dl-model-category}}
{\includegraphics[height=3.2cm]{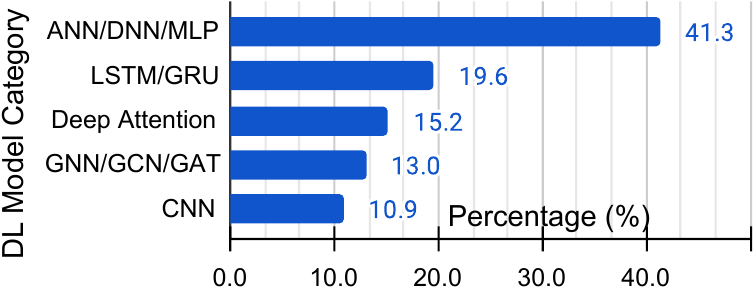}}
\hfill
\subcaptionbox{\footnotesize Distribution of hyper-architecture\label{fig:model-hyper-architecture}}
{\includegraphics[height=3.2cm]{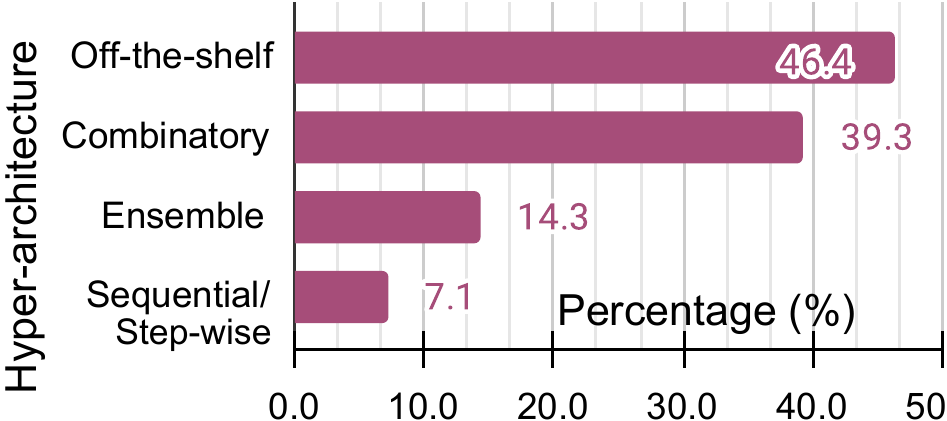}}
\caption{\label{fig:model-category-architecture}\textbf{The adopted DL model categories and hyper-architectures.}\newline The main DL model categories (backbones) are ANN-based (ANN/DNN/MLP), recurrent based (LSTM/GRU), deep attention \cite{vaswani2017attention}, graph based (GNN/GCN/GAT) and convolution based (CNN). These common models can be used either directly (off-the-shelf) or as basic building blocks to form different hyper-architectures: combinatory, ensemble or sequential.}
\end{figure}

Based on the experimental results from \cite{c23_kim2017does,c20_lee2018content,c22_yu2018prediction,c3_sharchilev2018web,w3_gastaud2019varying,c21_cheng2019success,j46_kim2020recommendation,c6_ghassemi2020automated,j41_yeh2020machine,j42_srinivasan2020ensemble,j44_kaminski2020predicting,c9_dellermann2021finding,c15_ferrati2021deep,c19_garkavenko2021valuation,c16_chen2021trend,j2_ross2021capitalvx,c1_zhang2021scalable,j26_bai2021startup,j37_kinne2021predicting,j43_shi2021leveraging,t1_stahl2021leveraging,t2_horn2021deep,w2_yin2021solving,w1_lyu2021graph,j16_allu2022predicting,j9_tang2022deep,b1_ang2022using,j47_wu2022estimating,w6_garkavenko2022you}, this section intends to provide insights to the choice of best-performing DL {\it model category} and {\it hyper-architecture}.
The model category refers to the basic backbone of the DL model, which include ANN/DNN/MLP\footnote{In this paper, ANN, DNN (deep neural network) and MLP (multi-layer perceptron) all refer to a neural network with at least two hidden layers, as introduced in Section~\ref{fig:dl-ann}.} \cite{b1_ang2022using,c15_ferrati2021deep,c16_chen2021trend,c22_yu2018prediction,c3_sharchilev2018web,c6_ghassemi2020automated,c9_dellermann2021finding,j2_ross2021capitalvx,j26_bai2021startup,j37_kinne2021predicting,j41_yeh2020machine,j42_srinivasan2020ensemble,j44_kaminski2020predicting,j47_wu2022estimating,t1_stahl2021leveraging,t2_horn2021deep,w2_yin2021solving,w6_garkavenko2022you}, LSTM (long short term memory) \cite{c15_ferrati2021deep,j16_allu2022predicting,j43_shi2021leveraging,j47_wu2022estimating,t2_horn2021deep,w1_lyu2021graph}, GRU (gated recurrent unit) \cite{c16_chen2021trend,c20_lee2018content,t1_stahl2021leveraging}, GNN (graph neural network) \cite{c1_zhang2021scalable,j47_wu2022estimating}, GCN (graph convolutional network) \cite{w3_gastaud2019varying,c16_chen2021trend}, GAT (graph attention network) \cite{w1_lyu2021graph,j47_wu2022estimating}, CNN (convolutional neural network) \cite{c15_ferrati2021deep,c21_cheng2019success,c23_kim2017does,j42_srinivasan2020ensemble,j43_shi2021leveraging} and deep attention \cite{vaswani2017attention,j9_tang2022deep,c16_chen2021trend,c20_lee2018content}. In Figure~\ref{fig:dl-model-category}, we group similar model categories (i.e. LSTM/GRU and GNN/GCN/GAT) together.
It can be seen that over 40\% of the surveyed papers adopt an ANN/DNN/MLP backbone due to its wide applicability to many data types.
LSTM/GRU almost dominates the cases when time-series are used.
Deep attention and graph based models (GNN/GCN/GAT) have a rising trend of adoption due to increasing introduction of text and graph input modalities.
Lastly, images and videos are relatively least used (cf. Figure~\ref{fig:data-modality-dist}), leading to only around 10\% adoption rate for CNN.

\begin{figure}[!t]
\centerline{\includegraphics[width=0.95\linewidth]{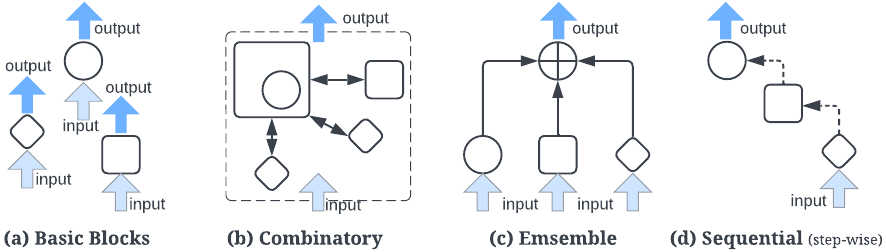}}
\vspace{-3pt}
\caption{\label{fig:hyper-architecture}\textbf{High-level illustration of different hyper-architectures.}\newline The off-the-shelf basic models are used as building blocks of different hyper-architectures in literature, where ``combinatory'' is the most complex and ``sequential'' is the simplest. } 
\end{figure}

Oftentimes, as shown in Figure~\ref{fig:model-hyper-architecture}, off-the-shelf models (i.e. specific implementation of a certain model category in Figure~\ref{fig:dl-model-category}) are used in research such as \cite{b1_ang2022using,c19_garkavenko2021valuation,c22_yu2018prediction,c6_ghassemi2020automated,j16_allu2022predicting,j26_bai2021startup,j37_kinne2021predicting,w6_garkavenko2022you,w1_lyu2021graph}. 
Some work manages to propose new architectures using existing (off-the-shelf) models as building blocks.
The particular way to ``glue'' these basic blocks together is termed {\it hyper-architecture}, which has three forms (cf. Figure~\ref{fig:hyper-architecture}): 
\begin{itemize}
\item Combinatory \cite{c1_zhang2021scalable,c15_ferrati2021deep,c16_chen2021trend,c20_lee2018content,c21_cheng2019success,c23_kim2017does,j43_shi2021leveraging,j47_wu2022estimating,t1_stahl2021leveraging,t2_horn2021deep}: the basic building blocks interact with (sometimes even embody) one another to achieve the same goal collaboratively;
\item Ensemble \cite{c3_sharchilev2018web,c9_dellermann2021finding,j2_ross2021capitalvx,j41_yeh2020machine}: each block still works independently, and their results will be aggregated to produce the final output.
\item Sequential (step-wise) \cite{j44_kaminski2020predicting,j46_kim2020recommendation}: a block can only start working when the previous block completes.
\end{itemize}

From the literature, it is almost impossible to draw a solid conclusion of which DL model works the best, 
because each work is often unique in sense of data sources, data modalities, data splitting, data processing, feature engineering, startup success criteria, evaluation metrics, and so on \cite{rj1_pasayat2020factors,rj2_song2008success}.
Unfortunately, the situation that each practitioner faces will continue to differ.

According to the {\bf No-Free-Lunch} ({\bf NFL}) theory \cite{wolpert1997no}, there is no such model that works the best in every situation: the model assumptions might fit for one situation yet fail to hold true for another. 
As a result, search for the optimal model for a particular setup (mainly the data and success definition) is important.
However, it is simply infeasible to attempt many model categories and hyper-architectures.
It is a common practice is to evaluate multiple models (cf. Section~\ref{sec:eval}) and pick the best-performing one, which is also reflected in the literature \cite{c23_kim2017does,c20_lee2018content,c22_yu2018prediction,c3_sharchilev2018web,w3_gastaud2019varying,c21_cheng2019success,j46_kim2020recommendation,c6_ghassemi2020automated,j41_yeh2020machine,j42_srinivasan2020ensemble,j44_kaminski2020predicting,c9_dellermann2021finding,c15_ferrati2021deep,c19_garkavenko2021valuation,c16_chen2021trend,j2_ross2021capitalvx,c1_zhang2021scalable,j26_bai2021startup,j37_kinne2021predicting,j43_shi2021leveraging,t1_stahl2021leveraging,t2_horn2021deep,w2_yin2021solving,w1_lyu2021graph,j16_allu2022predicting,j9_tang2022deep,b1_ang2022using,j47_wu2022estimating,w6_garkavenko2022you}.
Moreover, the data modality can sometimes imply the range of feasible DL model;
for example, LSTM/GRU is best suited to handle time-series data.

Finally, during this selection process, one should respect the principle of {\bf Occam's Razor} \cite{blumer1987occam}, which implies that one should prioritize the model with least complexity and best explainability.
This principle explains Figure~\ref{fig:model-category-architecture} where the simple off-the-shelf ANN/DNN/MLP model dominates.
Some work (e.g. \cite{w6_garkavenko2022you}) conclude that DL models do not necessarily yield better results than much simpler ML methods.
As a consequence, we recommend to start with testing simpler suitable model (even a pure random one); and then progressively increase the model complexity until the pre-allocated resources (e.g. man hour and GPU/CPU quota) are exhausted.

\section{Evaluate Model with Precision-First and Simulation Mindset}
\label{sec:eval}
\begin{figure}[!t]
\centerline{\includegraphics[width=0.9\linewidth]{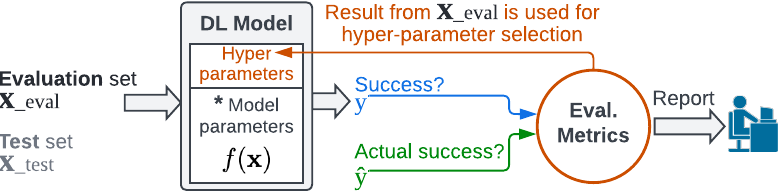}}
\caption{\label{fig:eval-test-process}\textbf{The Evaluation (hyper-parameter search) and Testing process.}\newline The evaluation/test data is fed into the trained model (hence the asterisk ``*'' beside the ``model parameters''), producing success prediction $y$. 
The evaluation metric estimates the prediction quality by comparing to the ground-truth labels $\hat{y}$.
The calculated metric values from the evaluation set ($\mathbf{x}_{\text{eval}}$) are used to guide selecting the best hyper-parameters.} 
\end{figure}

The decision of productizing any trained model is often made by looking at the evaluation results.
To achieve this goal, some {\it evaluation metrics}, as shown in Figure~\ref{fig:eval-test-process}, are employed to measure the quality of predictions (i.e. $y$) by comparing to the ground-truth labels (i.e. $\hat{y}$).
The metric values computed over the evaluation set (i.e. $\mathbf{x}_{\text{eval}}$) are used to determine which model (among many trained using different hyper-parameters) will be deployed for production eventually.
This process also fulfills the objective of searching for optimal hyper-parameters (i.e. hyper-parameter search).
It has been discussed previously, in Section~\ref{sec:dataset-split}, that the evaluation metrics for the test set (i.e. $\mathbf{x}_{\text{test}}$) are merely reported as an indication of the model's generalization capability\footnote{Generalization capability describes a model's ability to adapt properly to new, previously unseen data, drawn from the same distribution as the one used to train the model.}.

\begin{figure}[!t]
\centerline{\includegraphics[width=\linewidth]{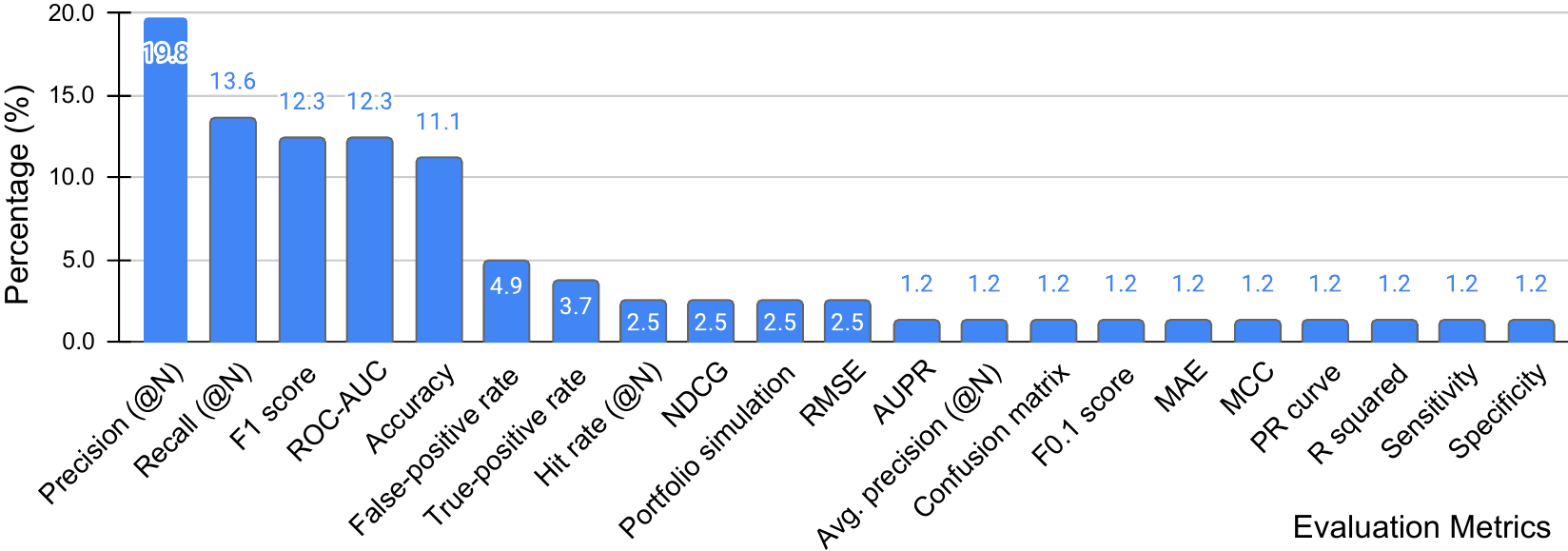}}
\vspace{-3pt}
\caption{\label{fig:eval-metrics-dist}\textbf{The distribution of adopted evaluation metrics in the surveyed DL literature.}\newline All metrics are quantitative indication of model performance. The notation ``@N'' implies the corresponding metric is calculated over a top-$N$ list. Precision, hit rate, and F0.1 score are popular metrics with a focus of high-precision. Portfolio simulation suited particularly well to the domain of startup success prediction, while others are general-purpose metrics for evaluating ML/DL models.} 
\end{figure}

The evaluation metrics adopted in the DL literature include (ordered by their occurrences as shown in Figure~\ref{fig:eval-metrics-dist}) {\it precision} \cite{c1_zhang2021scalable,c15_ferrati2021deep,c21_cheng2019success,j2_ross2021capitalvx,j37_kinne2021predicting,j42_srinivasan2020ensemble,j43_shi2021leveraging,j44_kaminski2020predicting,j9_tang2022deep,t1_stahl2021leveraging,t2_horn2021deep,w1_lyu2021graph,w2_yin2021solving,w3_gastaud2019varying,c3_sharchilev2018web}, {\it recall} \cite{c15_ferrati2021deep,c21_cheng2019success,j2_ross2021capitalvx,j37_kinne2021predicting,j42_srinivasan2020ensemble,j43_shi2021leveraging,j44_kaminski2020predicting,j9_tang2022deep,t2_horn2021deep,w2_yin2021solving,w3_gastaud2019varying}, {\it F1 score} \cite{c1_zhang2021scalable,c15_ferrati2021deep,c21_cheng2019success,j37_kinne2021predicting,j42_srinivasan2020ensemble,j43_shi2021leveraging,j44_kaminski2020predicting,j9_tang2022deep,w2_yin2021solving,w3_gastaud2019varying}, 
{\it ROC-AUC} ({\it area under the receiver operating characteristics}) \cite{c1_zhang2021scalable,c21_cheng2019success,c22_yu2018prediction,c3_sharchilev2018web,c6_ghassemi2020automated,j42_srinivasan2020ensemble,j43_shi2021leveraging,t1_stahl2021leveraging,t2_horn2021deep,w3_gastaud2019varying,w6_garkavenko2022you}, 
{\it accuracy} \cite{c1_zhang2021scalable,c20_lee2018content,c22_yu2018prediction,c9_dellermann2021finding,j2_ross2021capitalvx,j26_bai2021startup,j41_yeh2020machine,j42_srinivasan2020ensemble,j44_kaminski2020predicting,j9_tang2022deep}, 
{\it FPR} ({\it false-positive rate}) \cite{c6_ghassemi2020automated,t2_horn2021deep,w2_yin2021solving,w6_garkavenko2022you},
{\it TPR} ({\it true-positive rate}) \cite{c6_ghassemi2020automated,t2_horn2021deep,w6_garkavenko2022you},
{\it hit rate} \cite{j16_allu2022predicting,c16_chen2021trend},
{\it NDCG} ({\it normalized discounted cumulative gain}) \cite{c16_chen2021trend,j16_allu2022predicting},
{\it portfolio simulation} \cite{w2_yin2021solving,j2_ross2021capitalvx},
{\it RMSE} ({\it root mean square deviation}) \cite{c19_garkavenko2021valuation,j47_wu2022estimating},
{\it AUPR} ({\it area under the precision-recall curve}) \cite{c1_zhang2021scalable},
{\it average precision} \cite{w1_lyu2021graph},
{\it confusion matrix} \cite{j2_ross2021capitalvx},
{\it F0.1 score} \cite{c3_sharchilev2018web},
{\it MAE} ({\it mean absolute error}) \cite{j47_wu2022estimating},
{\it MCC} ({\it Matthews correlation coefficient}) \cite{c9_dellermann2021finding},
{\it PR curve} ({\it precision-recall curve}) \cite{t1_stahl2021leveraging},
{\it $R^2$ or ``R squared''} \cite{c19_garkavenko2021valuation}, and
{\it sensitivity/specificity} \cite{rj2_song2008success}.
To measure prediction quality, these evaluation metrics are formally similar to loss functions used in training [cf. Figure~\ref{fig:data-split-train-process}(b)], therefore some evaluation metrics (e.g. RMSE) can be used as loss functions too.

Most trained models are expected to act as a decision-support system for VC deal sourcing. 
Realistically, human investors are only able to assess a limited amount of startups.
Further, because of fund size limitation, investors can only fund a very small fraction of startups \cite{t1_stahl2021leveraging}. 
As a result, the {\bf evaluation metric should aim for high-precision (corresponding to high-certainty and low-recall)}\footnote{In the scope of VC deal sourcing, high-precision means the rate of ``correct'' prediction within the top-$N$ list (i.e. TPR: true-positive rate) should be high. According to the typical precision-recall curve, precision tends to be higher for smaller $N$; yet recall suffers from a small value of $N$.} \cite{c3_sharchilev2018web,t1_stahl2021leveraging}, which explains the popularity of {\it precision}, {\it true/false-positive rate}, {\it hit rate} and {\it F0.1 score} in Figure~\ref{fig:eval-metrics-dist}.

\begin{figure}[!t]
\centerline{\includegraphics[width=0.7\linewidth]{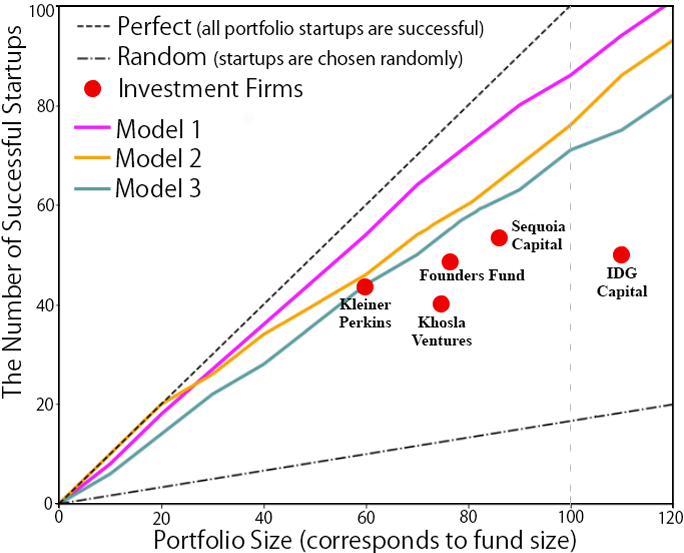}}
\vspace{-3pt}
\caption{\label{fig:portfolio-sim}\textbf{Evaluate model performance using simulated portfolios of different sizes.}\newline The trained DL model is used to form portfolios of different sizes $k\in \{20, 40, 60,80, 100, 120\}$ (x-axis); the number of eventually successful startups is plotted against the corresponding $k$, resulting in a model performance curve (i.e. the three colored curves). The perfect and random case are also plotted as straight lines showing the high and low performance boundary, respectively. The historical performance of real-world investment firms can be plotted as standalone points (red circles) for comparison. This chart is adapted from \cite{w2_yin2021solving}.} 
\end{figure}

There are four key questions to answer concerning any model trained to facilitate VC deal sourcing: 
{\bf Q1} What is the expected success ratio (or ROI) of the portfolio (with different sizes) constructed according to model predictions?
{\bf Q2} How will the model-driven portfolio perform in relation to the historical records of renowned investment firms?
{\bf Q3} Is the model significantly superior than a brainless random policy? 
{\bf Q4} How far does the model fall behind a theoretical perfect portfolio with 100\% success ratio?
Answering all questions simultaneously using any single general-purpose ML/DL metric is challenging and sometimes far-fetched.
To that end, some recent works \cite{j2_ross2021capitalvx,w2_yin2021solving} (though still far from a wide adoption according to Figure~\ref{fig:eval-metrics-dist}) have emerged proposing to {\bf evaluate via portfolio simulations}.
Recall that in Section~\ref{sec:dataset-split}, we recommended the investor-centric dataset split illustrated in Figure~\ref{fig:data-split-demo}. 
With that split, we make the trained models to predict the conditional success probability of each startup in evaluation/test set, using the last date of training period as the feature snapshot date.
Then, we construct an investment portfolio of size $k$ by selecting top-$k$ startups with the highest predicted probabilities.
As an indication of portfolio performance, we count the number of startups that eventually obtain a positive label.
The portfolio size $k$ should be varied,
so that we can plot one performance curve (the three colored curves in Figure~\ref{fig:portfolio-sim}) for each model.
To answer {\bf Q1}, a steeper curve corresponds to a better model.
The performance of a perfect model is a diagonal line, implying all portfolio startups will be successful.
To address {\bf Q2}, one just needs to measure the angular distance to diagonal.
The simplest possible model is a randomly policy, the performance of which is represented by the flattest straight line in Figure~\ref{fig:portfolio-sim}; 
the angular distance between this ``random'' line to any model's curve answers {\bf Q3}.
Finally, the historical fund performance of investment firms can be easily plotted as individual points, the vertical distances from which to models' curves give insights for {\bf Q4}.
Practically, the investment firms are more constrained than simulation: they can not invest in any startup due to many reasons like founders preference, portfolio conflict and investment mandate.
This constraint becomes more prominent when investors compete to invest in startups with great success potential.

\section{Resort to Model-Agnostic and Instance-Level Explainability}
\label{sec:explainability}

ANNs typically have thousands (often millions) of parameters, leading to extremely complicated nonlinear relationship between the input features/factors and output predictions.
As a result, DL models suffer from the criticism that they are mysterious black-boxes, as opposed to some white-box ML models like logistic regression.
However, explaining why a DL model comes up with certain predictions for startups (especially diverging predictions for comparable startups) is crucial for investment professionals \cite{j2_ross2021capitalvx}.
This requirement has been driving the application of Explainable AI (XAI) \cite{gade2019explainable} solutions in the field of startup success forecasting.
Practically, XAI helps to {\it increase trust in DL models} \cite{w2_yin2021solving}, 
{\it enable hypotheses-mining} \cite{w5_guerzoni2019survival},
{\it simplify model troubleshooting}, 
and {\it bust potential AI potholes} like biases.
Specifically, XAI aims to quantify the importance of every input feature/factor across all observations ({\bf global-level} explainability), or for one specific observation ({\bf instance-level} explainability) in the data.

For {\bf global-level} explainability, the contribution of any input feature is easily understood in the likes of regression models whose coefficients are directly associated with features: an increase of a feature by one unit increases the outcome by the amount of the corresponding coefficient.
But this approach does not apply to DL models, since there is no one-to-one relationships between model parameters and input features.
One solution is to simply aggregate the learned weights ({\it weights aggregation}) associated with a feature to approximate the impact of that feature \cite{j47_wu2022estimating}. 
For example, to measure the impact of feature $x_1$ in Figure~\ref{fig:dl-ann}, we may average all four weights connected to it.
This weight aggregation solution will fail quickly upon polarized weights or overly deep ANNs.
{\it Ablation analysis} \cite{w6_garkavenko2022you}, instead, generalizes better to various scenarios, where it investigates the model performance by removing certain input feature(s) to understand the contribution of the removed feature(s) to the overall system.
While feature importance here can be interpreted globally, it is not specific to any particular startup. 

\begin{figure}[!t]
\centerline{\includegraphics[width=0.9\linewidth]{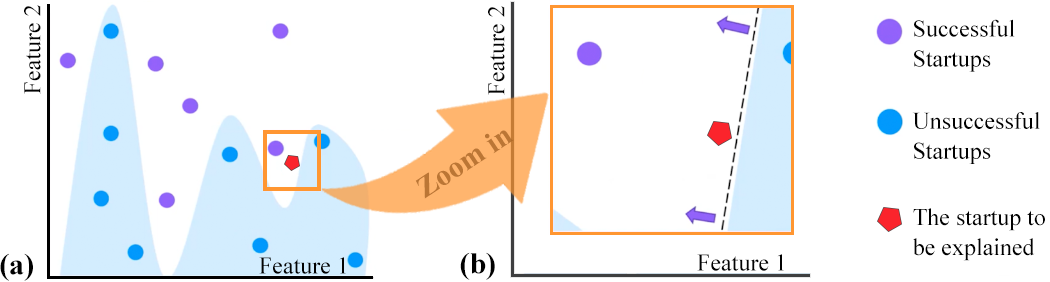}}
\vspace{-3pt}
\caption{\label{fig:lime}\textbf{LIME quantifies instance-level feature importance by examining local space.}\newline For simplicity, we assume each startup has merely two features (i.e. Feature 1 and 2).
The positive (successful) and negative (unsuccessful) startups are color coded differently, and separated by a squigly classification boundary.
To explain the prediction for a startup (represented by a red polygon), LIME examines the local space around this startup. 
In this example, the local separation boundary (linear approximation) turns out to be rather steep (i.e. a slope larger than 1.0), implying a much bigger impact from Feature 1.
}
\end{figure}

\begin{figure}[!t]
\centerline{\includegraphics[width=0.9\linewidth]{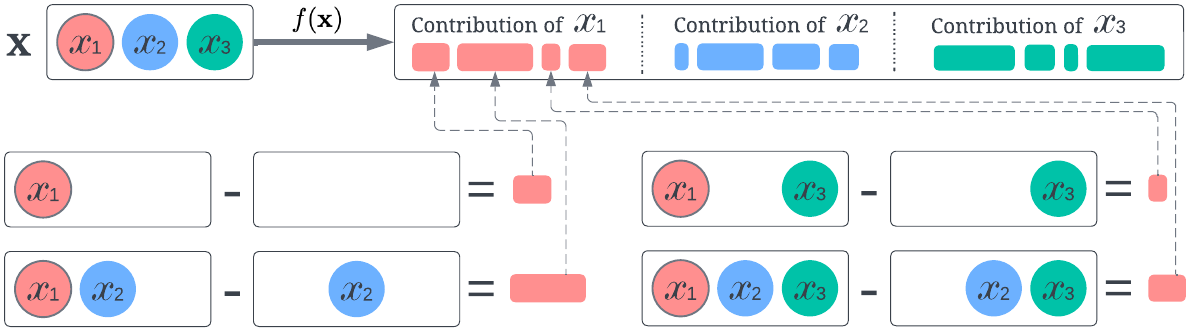}}
\vspace{-3pt}
\caption{\label{fig:shap}\textbf{A na\"ive illustration of how SHAP thinks about instance-level explainability.}\newline For simplicity, we assume that each startup is represented by a feature vector $\mathbf{x}$ with three features $x_1$, $x_2$ and $x_3$.
The model's prediction $f(\mathbf{x})$ can be viewed as a compound of contributions from $x_1$, $x_2$ and $x_3$.
To measure contribution of $x_1$, permutations of other features ($x_2$ and $x_3$) are created.
For each permutation, the output difference between the two cases (with and without $x_1$ in the input) are calculated.
Aggregating these differences for all permutations gives the final estimation for $x_1$. 
SHAP \cite{lundberg2017unified} proposes a way to efficiently approximate this permutation procedure so that it scales to high-dimension input.
}
\end{figure}

The {\bf instance-level} explainability is a little less straightforward due to the black-box nature of DL models; but it is perhaps the most valuable problem to address, since it provides fine-grained insight into why each individual prediction is made. 
It should also help explaining how the influence of a feature may vary with different observations that are fed to a DL model \cite{j2_ross2021capitalvx}. 
Given drastically different DL model categories and hyper-architectures in Figure~\ref{fig:model-category-architecture}, model-agnostic approaches are called for. Our survey suggests two popular methods:
\begin{itemize}
\item {\bf LIME} (Local Interpretable Model-agnostic Explanations) \cite{ribeiro2016should,j2_ross2021capitalvx}, as illustrated in Figure~\ref{fig:lime}, examines
the feature space local to an observation, 
and then applies a locally interpretable linear function $g(\mathbf{x})$ to approximate the model's black box function $f(\mathbf{x})$.
\item {\bf SHAP} (SHapley Additive exPlanations) \cite{j2_ross2021capitalvx,w2_yin2021solving,w3_gastaud2019varying,w6_garkavenko2022you},
as na\"ively shown in Figure~\ref{fig:shap}, measures the contribution of a certain feature $\mathbf{x}_1$ by permutating other features. In each permutation, the model will output one value with $\mathbf{x}_1$ (in the input), and output another without $\mathbf{x}_1$; the difference between these two values represents the contribution for that permutation.
Aggregating contributions from all permutations gives estimation for $\mathbf{x}_1$. 
The authors of SHAP \cite{lundberg2017unified} propose a way to approximate this permutation procedure so that it scales to large number of input features. 
\end{itemize}
SHAP is fast and deterministic, making it generally more attractive to researchers.
However, neither LIME nor SHAP is perfectly suited to explain extremely unstructured and high-dimensional modalities (e.g. texts and graphs in Figure~\ref{fig:data-modality-dist}).
This is why novel and tailored methods are being invented recently for explaining specific feature modalities such as text \cite{janizek2021explaining} and graph \cite{ying2019gnnexplainer}.

\section{Keep Humans in the Loop of Model Serving}
\label{sec:human-in-the-loop}
The end of model development is just the beginning of paying back the technical debts\footnote{Technical debt is a metaphor introduced by
Ward Cunningham in 1992 to help reason about the long term costs incurred by moving quickly in software engineering \cite{sculley2015hidden}.} \cite{sculley2015hidden} incurred from model productization$^{10}$,
during which humans (i.e.~the consumer of the model predictions) play a critical role in perceiving the prediction and providing feedback. 
Because of this, a few works \cite{c9_dellermann2021finding,j26_bai2021startup,t10_melnychuk2021approved} have attempted to shed lights on this topic.
We hereby briefly touch upon three most important aspects:

\begin{itemize}
\item {\bf Collective intelligence}: humans and DL models have advantage over one another and can make a unique contribution to informed investment decisions. 
Human decisions are intuitive, subjective, and sometimes bias-prone due to bounded rationality and mental shortcuts \cite{hoenig2015quality}.
However, they also have highly distilled domain knowledge that enables them to recognize and interpret very rare information. this leads to outcomes that are difficult to predict and would rather represent outliers in ML/DL models \cite{blattberg1990database}.
To that end, collective intelligence is suggested (by e.g. \cite{c9_dellermann2021finding,blohm2016rate,quinn2011human}) to complement the model prediction by assessing unknowable risk that can not be explained through data prior distribution.
\item {\bf Performance monitoring}: the model performance in production might be drastically different from evaluation/testing results (cf. Section~\ref{sec:eval}) due to the drift of input data and users' contexts.
As a result, monitoring users' feedback over prediction quality in a timely (if not real-time) manner is mandatory. For example, on weekly basis, each user (investment professional) is asked to assess a few startups (e.g. 5 to 10) that are highly likely to succeed according to the model.
Since precision is prioritized over other metrics (cf. Figure~\ref{fig:eval-metrics-dist}), we can monitor weekly precision as a gate keeper of model quality.
\item {\bf Model iteration}: degradation of the monitored metrics (e.g. precision) is expected at a certain time point, which is discussed in some industrial research such as \cite{cao2020debiasing}.
It is mostly caused by either drift of data distribution or change of user context.
While data drift could be corrected by retraining the model periodically using recent data, user context is much harder to compensate.
Here, user context refers to his/her way of defining the criteria of startup success or failure;
and the perturbance of such definition is hard to capture with additional heuristics as discussed in Section~\ref{sec:success_criteria}.
However, this problem may be relieved if we retrain the model using new assessments (both positive and negative) from users as additional labels.
\end{itemize}

\section{Conclusion}\label{s4}

Finding the rare unicorn startups is inherently challenging, hence often regarded as the holy grail for early-stage investors like Venture Capital (VC) firms.
Among the data driven approaches to help reaching that goal, deep learning (DL) has recently attracted more and more attention due to its superior model capacity and expressivity.
However, there has not been any comprehensive synthesis of the existing DL based research. 
This leaves many practitioners uninformed, and also vulnerable to some pitfalls hidden in nine key tasks -- problem scoping, success definition, data gathering, data processing, data split, model selection, model evaluation, model explanation, and model productization.
As a result, we carry out this literature synthesis on DL based methods with a practical lens.
The key statistics and learnings are presented from nine perspectives corresponding to the aforementioned nine tasks.
The authors' outlook of DL adoption in startup success prediction is three fold:
(1) more easy-to-use software tools will be developed to promote good practices and lower the barrier to entry;
(2) the majority of the available data is unlabeled and small scaled, hence more data/label efficient DL models will be proposed; 
(3) data privacy and model security will gain more emphasis in the coming years.

\section*{Acknowledgement}\label{ack}
We are particularly grateful to (alphabetically ordered by first names) 
Alex Patow (EQT Motherbrain), Anton Ask \r{A}str\"{o}m (EQT Ventures), Armin Catovic (EQT Motherbrain), Ashley Lundstr\"{o}m (EQT Ventures), Celine Xu (H\&M$\times$AI \& McKinsey), Daniel Wroblewski (CPPIB), Dhiana Deva (EQT Motherbrain), Drew McCornack (EQT Motherbrain), Fenni Kang (AntAlpha \& Barclays), Gustaf Halvardsson (KTH), Richard Stahl (EQT Growth \& EQT Motherbrain), Rockie Yang (Knock Data), Sofie Grant (EQT Ventures) and Wenbing Huang (Tsinghua University), whose input and review comments helped us to write and improve this paper.
We also thank the entire EQT Motherbrain team for their general support and interest of this research.

%% file: main.bbl
\begin{thebibliography}{100}

\bibitem{cao-etal-2023-using}
Lele Cao, Vilhelm von Ehrenheim, Sebastian Stan, Xiaoxue Li, and Alexandra
  Lutz.
\newblock Using deep learning to find the next unicorn: A practical synthesis
  on optimization target, feature selection, data split and evaluation
  strategy.
\newblock In {\em Proceedings of the IJCAI Joint Workshop on the 5th Financial
  Technology and Natural Language Processing (FinNLP) and the 2nd Multimodal AI
  for Financial Forecasting (Muffin)}, pages 63--73, Macao, 20 August 2023. ACL
  Anthology.

\bibitem{j8_santisteban2021critical}
Jos{\'e} Santisteban, David Mauricio, Orestes Cachay, et~al.
\newblock Critical success factors for technology-based startups.
\newblock {\em International Journal of Entrepreneurship and Small Business},
  42(4):397--421, 2021.

\bibitem{j22_skawinska2020success}
Eulalia Skawi{\'n}ska and Romuald~I Zalewski.
\newblock Success factors of startups in the {EU} - a comparative study.
\newblock {\em Sustainability}, 12(19):8200, 2020.

\bibitem{blank2013lean}
Steve Blank.
\newblock Why the lean start-up changes everything.
\newblock {\em Harvard business review}, 91(5):63--72, 2013.

\bibitem{acs2007entrepreneurship}
Zoltan~J Acs and Laszlo Szerb.
\newblock Entrepreneurship, economic growth and public policy.
\newblock {\em Small business economics}, 28(2):109--122, 2007.

\bibitem{c1_zhang2021scalable}
Shengming Zhang, Hao Zhong, Zixuan Yuan, and Hui Xiong.
\newblock Scalable heterogeneous graph neural networks for predicting
  high-potential early-stage startups.
\newblock In {\em ACM SIGKDD Conference on Knowledge Discovery and Data
  Mining}, pages 2202--2211, 2021.

\bibitem{marmer2011startup}
Max Marmer, Bjoern~Lasse Herrmann, Ertan Dogrultan, Ron Berman, Cuck Eesley,
  and Steve Blank.
\newblock Startup genome report extra: Premature scaling.
\newblock {\em Startup genome}, 10:1--56, 2011.

\bibitem{hyytinen2015does}
Ari Hyytinen, Mika Pajarinen, and Petri Rouvinen.
\newblock Does innovativeness reduce startup survival rates?
\newblock {\em Journal of business venturing}, 30(4):564--581, 2015.

\bibitem{j26_bai2021startup}
Sarah Bai and Yijun Zhao.
\newblock Startup investment decision support: Application of venture capital
  scorecards using machine learning approaches.
\newblock {\em Systems}, 9(3):55, 2021.

\bibitem{shane2012importance}
Scott Shane.
\newblock The importance of angel investing in financing the growth of
  entrepreneurial ventures.
\newblock {\em The Quarterly Journal of Finance}, 2(02):1250009, 2012.

\bibitem{t3_Unal2019Machine}
Cemre \"{U}nal and Ioana Ceasu.
\newblock A machine learning approach towards startup success prediction.
\newblock IRTG 1792 Discussion Paper 2019-022, Berlin, 2019.

\bibitem{teten2013lower}
David Teten, Adham AbdelFattah, Koen Bremer, and Gyorgy Buslig.
\newblock The lower-risk startup: how {V}enture {C}apitalists increase the odds
  of startup success.
\newblock {\em The Journal of Private Equity}, 16(2):7--19, 2013.

\bibitem{chernenko2021mutual}
Sergey Chernenko, Josh Lerner, and Yao Zeng.
\newblock Mutual funds as venture capitalists? evidence from unicorns.
\newblock {\em The Review of Financial Studies}, 34(5):2362--2410, 2021.

\bibitem{c9_dellermann2021finding}
Dominik Dellermann, Nikolaus Lipusch, Philipp Ebel, Karl~Michael Popp, and
  Jan~Marco Leimeister.
\newblock Finding the unicorn: Predicting early stage startup success through a
  hybrid intelligence method.
\newblock In {\em International Conference on Information Systems}, 2021.

\bibitem{cumming2010local}
Douglas Cumming and Na~Dai.
\newblock Local bias in venture capital investments.
\newblock {\em Journal of empirical finance}, 17(3):362--380, 2010.

\bibitem{lussier2001crossnational}
Robert~N Lussier and Sanja Pfeifer.
\newblock A crossnational prediction model for business success.
\newblock {\em Journal of small business management}, 39(3):228--239, 2001.

\bibitem{davila2003venture}
Antonio Davila, George Foster, and Mahendra Gupta.
\newblock Venture capital financing and the growth of startup firms.
\newblock {\em Journal of business venturing}, 18(6):689--708, 2003.

\bibitem{j19_hochberg2007whom}
Yael~V Hochberg, Alexander Ljungqvist, and Yang Lu.
\newblock Whom you know matters: Venture capital networks and investment
  performance.
\newblock {\em The Journal of Finance}, 62(1):251--301, 2007.

\bibitem{j34_nahata2008venture}
Rajarishi Nahata.
\newblock Venture capital reputation and investment performance.
\newblock {\em Journal of financial economics}, 90(2):127--151, 2008.

\bibitem{lussier2010three}
Robert~N Lussier and Claudia~E Halabi.
\newblock A three-country comparison of the business success versus failure
  prediction model.
\newblock {\em Journal of Small Business Management}, 48(3):360--377, 2010.

\bibitem{samila2011venture}
Sampsa Samila and Olav Sorenson.
\newblock Venture capital, entrepreneurship, and economic growth.
\newblock {\em The Review of Economics and Statistics}, 93(1):338--349, 2011.

\bibitem{puri2012life}
Manju Puri and Rebecca Zarutskie.
\newblock On the life cycle dynamics of venture-capital-and
  non-venture-capital-financed firms.
\newblock {\em The Journal of Finance}, 67(6):2247--2293, 2012.

\bibitem{nanda2013investment}
Ramana Nanda and Matthew Rhodes-Kropf.
\newblock Investment cycles and startup innovation.
\newblock {\em Journal of Financial Economics}, 110(2):403--418, 2013.

\bibitem{okrah2018exploring}
James Okrah, Alexander Nepp, and Ebenezer Agbozo.
\newblock Exploring the factors of startup success and growth.
\newblock {\em The Business and Management Review}, 9(3):229--237, 2018.

\bibitem{islam2018signaling}
Mazhar Islam, Adam Fremeth, and Alfred Marcus.
\newblock Signaling by early stage startups: {US} government research grants
  and venture capital funding.
\newblock {\em Journal of Business Venturing}, 33(1):35--51, 2018.

\bibitem{t7_saini2018picking}
Ajay Saini.
\newblock {\em Picking winners: A big data approach to evaluating startups and
  making venture capital investments}.
\newblock PhD thesis, Massachusetts Institute of Technology, 2018.

\bibitem{prohorovs2019startup}
Anatolijs Prohorovs, Julija Bistrova, and Daria Ten.
\newblock Startup success factors in the capital attraction stage: Founders’
  perspective.
\newblock {\em Journal of east-west business}, 25(1):26--51, 2019.

\bibitem{j33_malmstrom2020they}
Malin Malmstr{\"o}m, Aija Voitkane, Jeaneth Johansson, and Joakim Wincent.
\newblock What do they think and what do they say? gender bias, entrepreneurial
  attitude in writing and venture capitalists’ funding decisions.
\newblock {\em Journal of Business Venturing Insights}, 13:e00154, 2020.

\bibitem{gompers2020venture}
Paul~A Gompers, Will Gornall, Steven~N Kaplan, and Ilya~A Strebulaev.
\newblock How do venture capitalists make decisions?
\newblock {\em Journal of Financial Economics}, 135(1):169--190, 2020.

\bibitem{j30_kaiser2020value}
Ulrich Kaiser and Johan~M Kuhn.
\newblock The value of publicly available, textual and non-textual information
  for startup performance prediction.
\newblock {\em Journal of Business Venturing Insights}, 14:e00179, 2020.

\bibitem{rj1_pasayat2020factors}
Ajit~Kumar Pasayat, Bhaskar Bhowmick, and Ritik Roy.
\newblock Factors responsible for the success of a start-up: A meta-analytic
  approach.
\newblock {\em IEEE Transactions on Engineering Management}, 2020.

\bibitem{diaz2021econometric}
Carlos D{\'\i}az-Santamar{\'\i}a and Jacques Bulchand-Gidumal.
\newblock Econometric estimation of the factors that influence startup success.
\newblock {\em Sustainability}, 13(4):2242, 2021.

\bibitem{t10_melnychuk2021approved}
Anna Melnychuk.
\newblock {\em Startup Success Prediction: Example from The US}.
\newblock PhD thesis, Kyiv School of Economics, 2021.

\bibitem{williamson2002research}
Kirsty Williamson.
\newblock {\em Research methods for students, academics and professionals:
  Information management and systems}.
\newblock Elsevier, 2002.

\bibitem{w5_guerzoni2019survival}
Marco Guerzoni, Consuelo~R Nava, and Massimiliano Nuccio.
\newblock The survival of start-ups in time of crisis. a machine learning
  approach to measure innovation.
\newblock {\em arXiv preprint arXiv:1911.01073}, 2019.

\bibitem{c5_xiang2012supervised}
Guang Xiang, Zeyu Zheng, Miaomiao Wen, Jason Hong, Carolyn Rose, and Chao Liu.
\newblock A supervised approach to predict company acquisition with factual and
  topic features using profiles and news articles on techcrunch.
\newblock In {\em Proceedings of the International AAAI Conference on Web and
  Social Media}, volume~6, pages 607--610, 2012.

\bibitem{j36_rouhani2013erp}
S~Rouhani and Ahad~Zare Ravasan.
\newblock {ERP} success prediction: An artificial neural network approach.
\newblock {\em Scientia Iranica}, 20(3):992--1001, 2013.

\bibitem{j27_liang2016predicting}
Yuxian~Eugene Liang and Soe-Tsyr~Daphne Yuan.
\newblock Predicting investor funding behavior using crunchbase social network
  features.
\newblock {\em Internet research: Electronic networking applications and
  policy}, 26(1):74--100, 2016.

\bibitem{zhong2016or}
Hao Zhong, Chuanren Liu, Xinjiang Lu, and Hui Xiong.
\newblock To be or not to be friends: Exploiting social ties for venture
  investments.
\newblock In {\em International Conference on Data Mining}, pages 699--708,
  2016.

\bibitem{krishna2016predicting}
Amar Krishna, Ankit Agrawal, and Alok Choudhary.
\newblock Predicting the outcome of startups: less failure, more success.
\newblock In {\em International Conference on Data Mining Workshop}, pages
  798--805, 2016.

\bibitem{bohm2017business}
Markus B{\"o}hm, J{\"o}rg Weking, Frank Fortunat, Simon M{\"u}ller, Isabell
  Welpe, and Helmut Krcmar.
\newblock The business model {DNA}: Towards an approach for predicting business
  model success.
\newblock In {\em Internationale Tagung Wirtschaftsinformatik}, pages
  1006--1020, 2017.

\bibitem{zhong2018startup}
Hao Zhong, Chuanren Liu, Junwei Zhong, and Hui Xiong.
\newblock Which startup to invest in: a personalized portfolio strategy.
\newblock {\em Annals of Operations Research}, 263(1):339--360, 2018.

\bibitem{t5_bento2018predicting}
Francisco Ramadas da Silva~Ribeiro Bento.
\newblock {\em Predicting start-up success with machine learning}.
\newblock PhD thesis, Universidade NOVA de Lisboa, 2018.

\bibitem{arroyo2019assessment}
Javier Arroyo, Francesco Corea, Guillermo Jimenez-Diaz, and Juan~A
  Recio-Garcia.
\newblock Assessment of machine learning performance for decision support in
  venture capital investments.
\newblock {\em IEEE Access}, 7:124233--124243, 2019.

\bibitem{j15_shin2019network}
Sang~Yoon Shin.
\newblock Network advantage’s effect on exit performance: examining venture
  capital’s inter-organizational networks.
\newblock {\em International Entrepreneurship and Management Journal},
  15(1):21--42, 2019.

\bibitem{t6_unal2019searching}
Cemre {\"U}nal.
\newblock {\em Searching for a Unicorn: A Machine Learning Approach Towards
  Startup Success Prediction}.
\newblock PhD thesis, Humboldt University of Berlin, 2019.

\bibitem{li2020prediction}
Jinze Li.
\newblock Prediction of the success of startup companies based on support
  vector machine and random forset.
\newblock In {\em International Workshop on Artificial Intelligence and
  Education}, pages 5--11, 2020.

\bibitem{sadatrasoul2020hybrid}
Seyed~M Sadatrasoul, O~Ebadati, and R~Saedi.
\newblock A hybrid business success versus failure classification prediction
  model: A case of iranian accelerated start-ups.
\newblock {\em Journal of AI and Data Mining}, 8(2):279--287, 2020.

\bibitem{j17_bonaventura2020predicting}
Moreno Bonaventura, Valerio Ciotti, Pietro Panzarasa, Silvia Liverani, Lucas
  Lacasa, and Vito Latora.
\newblock Predicting success in the worldwide start-up network.
\newblock {\em Scientific Reports}, 10(1):1--6, 2020.

\bibitem{kipkogei2021tree}
Francis Kipkogei.
\newblock {\em Tree-based and Logistic Regression Models for Business Success
  Prediction in Rwanda}.
\newblock PhD thesis, University of Rwanda, 2021.

\bibitem{veloso2020predicting}
Felipe Veloso.
\newblock {\em Predicting Startup Success in {US}}.
\newblock PhD thesis, The University of North Carolina at Charlotte, 2020.

\bibitem{cavicchioli2021learning}
Maddalena Cavicchioli and Ulpiana Kocollari.
\newblock Learning from failure: Big data analysis for detecting the patterns
  of failure in innovative startups.
\newblock {\em Big Data}, 9(2):79--88, 2021.

\bibitem{zbikowski2021machine}
Kamil {\.Z}bikowski and Piotr Antosiuk.
\newblock A machine learning, bias-free approach for predicting business
  success using crunchbase data.
\newblock {\em Information Processing \& Management}, 58(4):102555, 2021.

\bibitem{t8_kamal2021modeling}
Adib Kamal and Kenan Sabani.
\newblock Modeling success factors for start-ups in western europe through a
  statistical learning approach.
\newblock Master's thesis, KTH Royal Institute of Technology, 2021.

\bibitem{singhal2022data}
Jaiesh Singhal, Chinmayi Rane, Yash Wadalkar, Mohit Joshi, and Amol Deshpande.
\newblock Data driven analysis for startup investments for venture capitalists.
\newblock In {\em International Conference for Advancement in Technology},
  pages 1--6, 2022.

\bibitem{bargagli2021supervised}
Falco~J Bargagli-Stoffi, Jan Niederreiter, and Massimo Riccaboni.
\newblock Supervised learning for the prediction of firm dynamics.
\newblock In {\em Data Science for Economics and Finance}, pages 19--41.
  Springer, Cham, 2021.

\bibitem{lecun2015deep}
Yann LeCun, Yoshua Bengio, and Geoffrey Hinton.
\newblock Deep learning.
\newblock {\em nature}, 521(7553):436--444, 2015.

\bibitem{goodfellow2016deep}
Ian Goodfellow, Yoshua Bengio, and Aaron Courville.
\newblock {\em Deep learning}.
\newblock MIT press, 2016.

\bibitem{raghu2017expressive}
Maithra Raghu, Ben Poole, Jon Kleinberg, Surya Ganguli, and Jascha
  Sohl-Dickstein.
\newblock On the expressive power of deep neural networks.
\newblock In {\em International Conference on Machine Learning}, pages
  2847--2854. PMLR, 2017.

\bibitem{hornik1989multilayer}
Kurt Hornik, Maxwell Stinchcombe, and Halbert White.
\newblock Multilayer feedforward networks are universal approximators.
\newblock {\em Neural Networks}, 2(5):359--366, 1989.

\bibitem{c23_kim2017does}
Jongho Kim and Jiyong Park.
\newblock Does facial expression matter even online? an empirical analysis of
  facial expression of emotion and crowdfunding success.
\newblock In {\em International Conference on Information Systems}, 2017.

\bibitem{c20_lee2018content}
SeungHun Lee, KangHee Lee, and Hyun-chul Kim.
\newblock Content-based success prediction of crowdfunding campaigns: A deep
  learning approach.
\newblock In {\em Companion of the ACM Conference on Computer Supported
  Cooperative Work and Social Computing}, pages 193--196, 2018.

\bibitem{c22_yu2018prediction}
Pi-Fen Yu, Fu-Ming Huang, Chuan Yang, Yu-Hsin Liu, Zi-Yi Li, and Cheng-Hung
  Tsai.
\newblock Prediction of crowdfunding project success with deep learning.
\newblock In {\em International Conference on E-Business Engineering}, pages
  1--8. IEEE, 2018.

\bibitem{c3_sharchilev2018web}
Boris Sharchilev, Michael Roizner, Andrey Rumyantsev, Denis Ozornin, Pavel
  Serdyukov, and Maarten de~Rijke.
\newblock Web-based startup success prediction.
\newblock In {\em International Conference on Information and Knowledge
  Management}, pages 2283--2291, 2018.

\bibitem{w3_gastaud2019varying}
Clement Gastaud, Theophile Carniel, and Jean-Michel Dalle.
\newblock The varying importance of extrinsic factors in the success of startup
  fundraising: competition at early-stage and networks at growth-stage.
\newblock {\em arXiv preprint arXiv:1906.03210}, 2019.

\bibitem{c21_cheng2019success}
Chaoran Cheng, Fei Tan, Xiurui Hou, and Zhi Wei.
\newblock Success prediction on crowdfunding with multimodal deep learning.
\newblock In {\em International Joint Conference on Artificial Intelligence},
  pages 2158--2164, 2019.

\bibitem{j46_kim2020recommendation}
Hyoung~Jun Kim, Tae San~Kim, and So~Young Sohn.
\newblock Recommendation of startups as technology cooperation candidates from
  the perspectives of similarity and potential: A deep learning approach.
\newblock {\em Decision Support Systems}, 130:113229, 2020.

\bibitem{c6_ghassemi2020automated}
M~Ghassemi, C~Song, and T~Alhanai.
\newblock The automated venture capitalist: Data and methods to predict the
  fate of startup ventures.
\newblock In {\em AAAI Workshop on Knowledge Discovery from Unstructured Data
  in Financial Services}, 2020.

\bibitem{j41_yeh2020machine}
Jen-Yin Yeh and Chi-Hua Chen.
\newblock A machine learning approach to predict the success of crowdfunding
  fintech project.
\newblock {\em Journal of Enterprise Information Management}, ahead-of-print,
  2020.

\bibitem{j42_srinivasan2020ensemble}
Arvind Srinivasan et~al.
\newblock An ensemble deep learning approach to explore the impact of
  enticement, engagement and experience in reward based crowdfunding.
\newblock Working paper, Department of Computer Science and Engineering, SRM
  Institute of Science and Technology, 2020.

\bibitem{j44_kaminski2020predicting}
Jermain~C Kaminski and Christian Hopp.
\newblock Predicting outcomes in crowdfunding campaigns with textual, visual,
  and linguistic signals.
\newblock {\em Small Business Economics}, 55(3):627--649, 2020.

\bibitem{c15_ferrati2021deep}
Francesco Ferrati, Haiquan Chen, and Moreno Muffatto.
\newblock A deep learning model for startups evaluation using time series
  analysis.
\newblock In {\em European Conference on Innovation and Entrepreneurship}, page
  311. Academic Conferences limited, 2021.

\bibitem{c19_garkavenko2021valuation}
Mariia Garkavenko, Hamid Mirisaee, Eric Gaussier, Agn{\`e}s Guerraz, and
  C{\'e}dric Lagnier.
\newblock Valuation of startups: A machine learning perspective.
\newblock In {\em European Conference on Information Retrieval}, pages
  176--189. Springer, 2021.

\bibitem{c16_chen2021trend}
Miao Chen, Chao Wang, Chuan Qin, Tong Xu, Jianhui Ma, Enhong Chen, and Hui
  Xiong.
\newblock A trend-aware investment target recommendation system with
  heterogeneous graph.
\newblock In {\em International Joint Conference on Neural Networks}, pages
  1--8. IEEE, 2021.

\bibitem{j2_ross2021capitalvx}
Greg Ross, Sanjiv Das, Daniel Sciro, and Hussain Raza.
\newblock Capitalvx: A machine learning model for startup selection and exit
  prediction.
\newblock {\em The Journal of Finance and Data Science}, 7:94--114, 2021.

\bibitem{j37_kinne2021predicting}
Jan Kinne and David Lenz.
\newblock Predicting innovative firms using web mining and deep learning.
\newblock {\em PloS One}, 16(4):e0249071, 2021.

\bibitem{j43_shi2021leveraging}
Jiatong Shi, Kunlin Yang, Wei Xu, and Mingming Wang.
\newblock Leveraging deep learning with audio analytics to predict the success
  of crowdfunding projects.
\newblock {\em The Journal of Supercomputing}, 77(7):7833--7853, 2021.

\bibitem{t1_stahl2021leveraging}
Richard Hermann~Anselmo Stahl.
\newblock Leveraging time-series signals for multi-stage startup success
  prediction.
\newblock Master's thesis, Swiss Federal Institute of Technology (ETH Zurich),
  2021.

\bibitem{t2_horn2021deep}
Sonja Horn.
\newblock Deep learning models as decision support in venture capital
  investments: Temporal representations in employee growth forecasting of
  startup companies.
\newblock Master's thesis, KTH Royal Institute of Technology, 2021.

\bibitem{w2_yin2021solving}
Dafei Yin, Jing Li, and Gaosheng Wu.
\newblock Solving the data sparsity problem in predicting the success of the
  startups with machine learning methods.
\newblock {\em arXiv preprint arXiv:2112.07985}, 2021.

\bibitem{w1_lyu2021graph}
Shiwei Lyu, Shuai Ling, Kaihao Guo, Haipeng Zhang, Kunpeng Zhang, Suting Hong,
  Qing Ke, and Jinjie Gu.
\newblock Graph neural network based vc investment success prediction.
\newblock {\em arXiv preprint arXiv:2105.11537}, 2021.

\bibitem{j16_allu2022predicting}
Ramakrishna Allu and Venkata Nageswara~Rao Padmanabhuni.
\newblock Predicting the success rate of a start-up using lstm with a swish
  activation function.
\newblock {\em Journal of Control and Decision}, 9(3):355--363, 2022.

\bibitem{j9_tang2022deep}
Zhe Tang, Yi~Yang, Wen Li, Defu Lian, and Lixin Duan.
\newblock Deep cross-attention network for crowdfunding success prediction.
\newblock {\em IEEE Transactions on Multimedia}, ahead-of-print, 2022.

\bibitem{b1_ang2022using}
Yu~Qian Ang, Andrew Chia, and Soroush Saghafian.
\newblock Using machine learning to demystify startups’ funding, post-money
  valuation, and success.
\newblock In {\em Innovative Technology at the Interface of Finance and
  Operations}, pages 271--296. Springer, 2022.

\bibitem{j47_wu2022estimating}
Likang Wu, Zhi Li, Hongke Zhao, Qi~Liu, and Enhong Chen.
\newblock Estimating fund-raising performance for start-up projects from a
  market graph perspective.
\newblock {\em Pattern Recognition}, 121:108204, 2022.

\bibitem{w6_garkavenko2022you}
Mariia Garkavenko, Eric Gaussier, Hamid Mirisaee, C{\'e}dric Lagnier, and
  Agn{\`e}s Guerraz.
\newblock Where do you want to invest? predicting startup funding from freely,
  publicly available web information.
\newblock {\em arXiv preprint arXiv:2204.06479}, 2022.

\bibitem{b2_corea2019ai}
Francesco Corea.
\newblock {AI} and venture capital.
\newblock In {\em An introduction to data}, pages 101--110. Springer, 2019.

\bibitem{baxter1997bayesian}
Jonathan Baxter.
\newblock A bayesian/information theoretic model of learning to learn via
  multiple task sampling.
\newblock {\em Machine Learning}, 28(1):7--39, 1997.

\bibitem{baer2014gold}
John Baer and Sharon~S McKool.
\newblock The gold standard for assessing creativity.
\newblock {\em International Journal of Quality Assurance in Engineering and
  Technology Education (IJQAETE)}, 3(1):81--93, 2014.

\bibitem{ewens2018founder}
Michael Ewens and Matt Marx.
\newblock Founder replacement and startup performance.
\newblock {\em The Review of Financial Studies}, 31(4):1532--1565, 2018.

\bibitem{c2_pasayat2021evolutionary}
Ajit~Kumar Pasayat and Bhaskar Bhowmick.
\newblock An evolutionary algorithm-based framework for determining crucial
  features contributing to the success of a start-up.
\newblock In {\em IEEE Technology and Engineering Management Conference-Europe
  (TEMSCON-EUR)}, pages 1--6. IEEE, 2021.

\bibitem{rj2_song2008success}
Michael Song, Ksenia Podoynitsyna, Hans Van Der~Bij, and Johannes~IM Halman.
\newblock Success factors in new ventures: A meta-analysis.
\newblock {\em Journal of Product Innovation Management}, 25(1):7--27, 2008.

\bibitem{cao-etal-2022-sire}
Lele Cao, Sonja Horn, Vilhelm von Ehrenheim, Richard~Anselmo Stahl, and Henrik
  Landgren.
\newblock Simulation-informed revenue extrapolation with confidence estimate
  for scaleup companies using scarce time series data.
\newblock In {\em Proceedings of the 31st ACM International Conference on
  Information and Knowledge Management (CIKM ’22), October 17–21, 2022,
  Atlanta, GA, USA}, page 12 pages, New York, NY, USA, October 2022.
  Association for Computing Machinery (ACM).

\bibitem{johnson2019survey}
Justin~M Johnson and Taghi~M Khoshgoftaar.
\newblock Survey on deep learning with class imbalance.
\newblock {\em Journal of Big Data}, 6(1):1--54, 2019.

\bibitem{chawla2002smote}
Nitesh~V Chawla, Kevin~W Bowyer, Lawrence~O Hall, and W~Philip Kegelmeyer.
\newblock Smote: synthetic minority over-sampling technique.
\newblock {\em Journal of Artificial Intelligence Research}, 16:321--357, 2002.

\bibitem{he2008adasyn}
Haibo He, Yang Bai, Edwardo~A Garcia, and Shutao Li.
\newblock Adasyn: Adaptive synthetic sampling approach for imbalanced learning.
\newblock In {\em IEEE International Joint Conference on Neural Networks (IEEE
  World Congress on Computational Intelligence)}, pages 1322--1328. IEEE, 2008.

\bibitem{kiryo2017positive}
Ryuichi Kiryo, Gang Niu, Marthinus~C du~Plessis, and Masashi Sugiyama.
\newblock Positive-unlabeled learning with non-negative risk estimator.
\newblock In {\em Proceedings of the 31st International Conference on Neural
  Information Processing Systems (NeurIPS)}, pages 1674--1684, 2017.

\bibitem{wei_zou_2019_eda}
Jason Wei and Kai Zou.
\newblock {EDA}: Easy data augmentation techniques for boosting performance on
  text classification tasks.
\newblock In {\em Proceedings of the 2019 Conference on Empirical Methods in
  Natural Language Processing and the 9th International Joint Conference on
  Natural Language Processing (EMNLP-IJCNLP)}, pages 6382--6388, Hong Kong,
  China, November 2019. Association for Computational Linguistics.

\bibitem{yao2017recent}
Jin-ge Yao, Xiaojun Wan, and Jianguo Xiao.
\newblock Recent advances in document summarization.
\newblock {\em Knowledge and Information Systems}, 53(2):297--336, 2017.

\bibitem{jager2021benchmark}
Sebastian J{\"a}ger, Arndt Allhorn, and Felix Bie{\ss}mann.
\newblock A benchmark for data imputation methods.
\newblock {\em Frontiers in Big Data}, page~48, 2021.

\bibitem{mazumder2010spectral}
Rahul Mazumder, Trevor Hastie, and Robert Tibshirani.
\newblock Spectral regularization algorithms for learning large incomplete
  matrices.
\newblock {\em The Journal of Machine Learning Research}, 11:2287--2322, 2010.

\bibitem{vaswani2017attention}
Ashish Vaswani, Noam Shazeer, Niki Parmar, Jakob Uszkoreit, Llion Jones,
  Aidan~N Gomez, {\L}ukasz Kaiser, and Illia Polosukhin.
\newblock Attention is all you need.
\newblock {\em Advances in neural information processing systems}, 30, 2017.

\bibitem{wolpert1997no}
David~H Wolpert and William~G Macready.
\newblock No free lunch theorems for optimization.
\newblock {\em IEEE Transactions on Evolutionary Computation}, 1(1):67--82,
  1997.

\bibitem{blumer1987occam}
Anselm Blumer, Andrzej Ehrenfeucht, David Haussler, and Manfred~K Warmuth.
\newblock Occam's razor.
\newblock {\em Information Processing Letters}, 24(6):377--380, 1987.

\bibitem{gade2019explainable}
Krishna Gade, Sahin~Cem Geyik, Krishnaram Kenthapadi, Varun Mithal, and Ankur
  Taly.
\newblock Explainable {AI} in industry.
\newblock In {\em ACM SIGKDD International Conference on Knowledge Discovery
  and Data Mining}, pages 3203--3204, 2019.

\bibitem{ribeiro2016should}
Marco~Tulio Ribeiro, Sameer Singh, and Carlos Guestrin.
\newblock ``{W}hy should {I} trust you?'' {E}xplaining the predictions of any
  classifier.
\newblock In {\em ACM SIGKDD International Conference on Knowledge Discovery
  and Data Mining}, pages 1135--1144, 2016.

\bibitem{lundberg2017unified}
Scott~M Lundberg and Su-In Lee.
\newblock A unified approach to interpreting model predictions.
\newblock In {\em International Conference on Neural Information Processing
  Systems}, pages 4768--4777, 2017.

\bibitem{janizek2021explaining}
Joseph~D Janizek, Pascal Sturmfels, and Su-In Lee.
\newblock Explaining explanations: Axiomatic feature interactions for deep
  networks.
\newblock {\em Journal of Machine Learning Research}, 22:1--54, 2021.

\bibitem{ying2019gnnexplainer}
Zhitao Ying, Dylan Bourgeois, Jiaxuan You, Marinka Zitnik, and Jure Leskovec.
\newblock {GNN}explainer: {G}enerating explanations for graph neural networks.
\newblock In {\em International Conference on Neural Information Processing
  Systems}, volume~32, pages 9240--9251, 2019.

\bibitem{sculley2015hidden}
D~Sculley, Gary Holt, Daniel Golovin, Eugene Davydov, Todd Phillips, Dietmar
  Ebner, Vinay Chaudhary, Michael Young, Jean-Francois Crespo, and Dan
  Dennison.
\newblock Hidden technical debt in machine learning systems.
\newblock In {\em International Conference on Neural Information Processing
  Systems}, volume~2, pages 2503--2511, 2015.

\bibitem{hoenig2015quality}
Daniel Hoenig and Joachim Henkel.
\newblock Quality signals? the role of patents, alliances, and team experience
  in venture capital financing.
\newblock {\em Research Policy}, 44(5):1049--1064, 2015.

\bibitem{blattberg1990database}
Robert~C Blattberg and Stephen~J Hoch.
\newblock Database models and managerial intuition: 50\% model+ 50\% manager.
\newblock {\em Management Science}, 36(8):887--899, 1990.

\bibitem{blohm2016rate}
Ivo Blohm, Christoph Riedl, Johann F{\"u}ller, and Jan~Marco Leimeister.
\newblock Rate or trade? identifying winning ideas in open idea sourcing.
\newblock {\em Information Systems Research}, 27(1):27--48, 2016.

\bibitem{quinn2011human}
Alexander~J Quinn and Benjamin~B Bederson.
\newblock Human computation: a survey and taxonomy of a growing field.
\newblock In {\em SIGCHI Conference on Human Factors in Computing Systems},
  pages 1403--1412, 2011.

\bibitem{cao2020debiasing}
Lele Cao, Sahar Asadi, Matteo Biasielli, and Michael Sj{\"o}berg.
\newblock Debiasing few-shot recommendation in mobile games.
\newblock In {\em The ORSUM (Online Recommender Systems and User Modeling)
  Workshop in conjunction with the ACM Conference on Recommender Systems
  (RecSys)}, 2020.

\end{thebibliography}
